\documentclass[a4paper,fleqn]{cas-dc}
\usepackage[numbers,sort&compress]{natbib}
\usepackage{amsmath,amsfonts}
\usepackage{array}
\usepackage{amssymb} 
\usepackage{textcomp}
\usepackage{stfloats}
\usepackage{verbatim}
\usepackage{graphicx}
\usepackage{acronym}
\usepackage{caption, subcaption}
 
\usepackage{float}
\usepackage{subfloat}
\usepackage{comment}
\usepackage{soul}
\usepackage{color}
\usepackage{mathtools}
\usepackage{multicol}
\def\tsc#1{\csdef{#1}{\textsc{\lowercase{#1}}\xspace}}
\tsc{WGM}
\tsc{QE}
\begin{document}
\let \WriteBookmarks \relax
\def \textpagefraction {.001}
\let\printorcid\relax
\shorttitle{General Scintillation for Gaussian Beam Propagating through Oceanic Turbulence and UWOC System Performance Evaluation} 
\shortauthors{Yuxuan~Li et al}
\title [mode = title]{General Scintillation for Gaussian Beam Propagating through Oceanic Turbulence and UWOC System Performance Evaluation}  
\author[1]{Yuxuan~Li}
\ead{lyuxuan@stu.xidian.edu.cn}
\author[1]{Xiang~Yi}
\cormark[1]
\ead{yixiang@xidian.edu.cn}
\cortext[1]{Corresponding author:yixiang@xidian.edu.cn}
\author[1]{Xinyue~Tao}
\ead{xyt@stu.xidian.edu.cn}
\author[2]{Ata~Yalçın}
\ead{ylcnata@gmail.com}
\author[3]{Mingjian~Cheng}
\author[4]{Lu~Zhang}
\address[1]{School of Communications Engineering, Xidian University, Xi’an, Shaanxi 710071, China}
\address[2]{Department of Electrical and Electronics Engineering, OSTIM Technical University, 06374 Yenimahalle, Ankara, Turkey}
\address[3]{School of Physics, Xidian University, Xi’an,
Shaanxi 710071, China}
\address[4]{School of Optoelectronic Engineering, Xidian University, Xi’an, Shaanxi 710071, China}
\begin{abstract}
In this paper, we derive a general and exact closed-form expression of scintillation index (SI) for a Gaussian beam propagating through weak oceanic turbulence, based on the general oceanic turbulence optical power spectrum (OTOPS) and the Rytov theory. Our universal expression not only includes existing Rytov variances but also accounts for actual cases where the Kolmogorov microscale is non-zero. The correctness and accuracy of our derivation are verified through comparison with the published work under identical conditions. By utilizing our derived expressions, we analyze the impact of various beam, propagation and oceanic turbulence parameters on both SI and bit error rate (BER) performance of underwater wireless optical communication (UWOC) systems. Numerical results demonstrate that the relationship between the Kolmogorov microscale and SI is nonlinear. Additionally, considering that certain oceanic turbulence parameters are related to depth, we use temperature and salinity data from Argo buoy deployed in real oceans to investigate the dependence of SI on depth. Our findings will contribute to the design and optimization of UWOC systems.
\end{abstract}
\begin{keywords}
Scintillation index\sep Underwater wireless optical communication\sep Optical wireless channel\sep
\end{keywords}
\maketitle
\section{Introduction} \label{sec:introduction}
With the increasing demands for underwater applications, such as environmental monitoring, marine biology, tracking seismic activities, surveillance and navigation, the underwater data transmission technologies that can achieve high efficiency and reliability have witnessed growing interest in recent years. Among them, underwater wireless optical communication (UWOC) has received specific attention due to its advantages of high data rate, low delay and high security \cite{ref01,ref02,ref03}. However, in addition to the absorption and scattering of seawater, the performance of UWOC systems can also be hampered by oceanic turbulence, which represents the refractive-index random fluctuations caused by temperature and salinity variations. The presence of turbulence can lead to the random fluctuations of the optical signal irradiance intensity, causing it to fade below the threshold, ultimately resulting in communication interrupt. Such an effect needs to be quantified by the scintillation index (SI).\\
\indent Several works have already been done by theoretically prediction of the SI based on the oceanic turbulence optical power spectrum (OTOPS) and the Rytov approximation approach. OTOPS well describes the statistical characteristics of turbulent random field in spatial-frequency domain \cite{ref04}. In terms of the principle of fluid dynamics, Hill proposed four optical power spectrum models for scalar (such as temperature, humidity, or salinity) fluctuations. Among them, only Model 1 has a closed-form solution, which facilitates the mathematical derivation. But Model 4 shows the most accuracy as fitting the experimental data \cite{ref05,ref06}. Nikishov's power spectrum model developed from Hill Model 1, as the earliest OTOPS, has a closed expression and contains not only temperature and salinity spectrums but also thermohaline coupling spectrum \cite{ref07}. Based on \cite{ref07}, the SI of plane wave, spherical wave, and Gaussian beam under the weak turbulence conditions were studied. But due to the mathematical complicacy, the analysis was restricted to numerical calculation \cite{ref08,ref09}. Also using Nikishov's power spectrum model, the analytical expressions of SI for plane and spherical waves under limited receiving aperture were derived \cite{ref10}. The associated bit-error-rate (BER) performance of UWOC systems was examined. The obtained SI and BER are significantly lower than those calculated for point receiving. While infinite plane and spherical waves made the mathematical process of SI simpler, they are too ideal to describe the finite communication and propagation scenario. It is more meaningful to use Gaussian beam which is commonly used to model the transmitted optical field of laser. The analytical expression of SI for a Gaussian beam was firstly deduced in \cite{ref11}. The resulting formula yields the SI values that are identical to the numerical results found in \cite{ref09}. Then, this SI was used to evaluated the BER performance of UWOC systems \cite{ref12}. These works were further carried out for limited receiving aperture \cite{ref13}.\\
\indent Nikishov's power spectrum model deviates the experiment results significantly in the viscous-convective portion of the spectrum, despite having a tractable expression \cite{ref14,ref15}. To overcome its limitations, Yi's power spectrum model based on Hill Model 4 was proposed by using Kraichnan spectrum instead of Batchelor spectrum to match the experimental data of viscous-convective range. Based on this proposed power spectrum model and Rytov theory, the numerical calculation of SI for plane and spherical waves were performed \cite{ref16}. The results showed that Nikishov's power spectrum model obviously underestimated the SI values. However, Yi's power spectrum model was only applicable for fixed Prandtl/Schmidt numbers, and its complicated mathematical form made it difficult to obtain a closed expression of SI. In this context, a wide range of Prandtl/Schmidt number power spectrum model with a compact mathematical form was proposed by analytically solving Hill Model 4, the SI of spherical wave was also derived \cite{ref17}. Subsequently, the power spectrum model in \cite{ref17} was extended to a more general model, namely, the general OTOPS, whose parameters were correlated with the temperature, salinity of seawater and fluid dynamics parameters, further enhancing the practicability of this spectrum \cite{ref18}.\\
\indent Using the above-mentioned general OTOPS \cite{ref18}, the SI expressions of various optical beams were derived and the associated UWOC performance analysis were also carried out. The Rytov variances of plane and spherical waves were derived in \cite{ref19} under weak turbulence conditions. The SI for these infinite optical waves in moderate-to-strong oceanic turbulence were derived by means of extended Rytov theory \cite{ref20}. The obtained SI was used to investigate the BER performance of UWOC \cite{ref20}. For limited communication distances and finite transceiver aperture, it is more meaningful to study the SI of bounded beam such as Gaussian beam for the design of actual UWOC system. However, the complicated mathematically form of the general OTOPS and Gaussian beam make it difficult to get a tractable expression. Recently, several studies have been performed to obtain computable expressions through some approximations, simplifications and assumptions. The approximate expression of the Rytov variance was derived for the point receiving by assuming Kolmogorov microscale tends to zero \cite{ref21}. Besides this assumption it was used eliminating one positive and one negative components in the derivation, which simplified the mathematical process, but leading to less accuracy. Following the similar procedure, the approximate SI was obtained for a finite receiving aperture \cite{ref22}. A simplified expression of SI for Gaussian beam was derived with a certain degree of approximation \cite{ref23}. Although the expression contains Kolmogorov microscale parameter, its results were just efficient for short-to-medium distances and specific range of receiving aperture. To sum up, there has not been a general and exact expression for the SI of Gaussian beam in oceanic turbulence, to the best of our knowledge.\\
\indent Motivated by this, we derive a general and exact closed-form analytical expression for the SI of Gaussian beam by employing the general OTOPS and the Rytov approach. Our results not only include the previous results in \cite{ref19,ref21} as special cases, but also extend to general case that Kolmogorov microscale is non-zero, considering impacts of all oceanic turbulence parameters. The accuracy of our derivation is also verified by the comparison with existing approximate results. In addition, using temperature and salinity data from Argo buoy deployed in real oceans, we investigate the depth dependence of SI in the vertical direction. We further use the derived SI to evaluate the BER performance of UWOC system under weak oceanic turbulence.\\
\indent The rest of this paper is organized as follows. The expression of general on-axis scintillation index for Gaussian beam is derived in Section \ref{sec:derivation}. The analytical derivation of BER performance is illustrated in Section \ref{sec:BER}. The numerical results and related analysis are presented in Section \ref{sec:results} and the conclusions about our works are drawn in Section \ref{sec:conclusion}.
\section{General on-axis Scintillation Index of Gaussian Beam} \label{sec:derivation}
\subsection{General OTOPS Model}
The general OTOPS model can be expressed as \cite{ref18}
\begin{equation} \label{equ:01}
    \Phi_n(\kappa,\left \langle T \right \rangle,\left \langle S \right \rangle,\lambda)
= A^2\Phi_T(\kappa)+B^2\Phi_S(\kappa)+2AB\Phi_{TS}(\kappa) 
\end{equation}
where $\kappa$ is the spatial frequency, $A$ and $B$ are linear coefficients related to the average temperature $\left \langle T \right \rangle$, average salinity concentration $\left \langle S \right \rangle$ and the wavelength $\lambda$. They are defined as
\begin{equation} \label{equ:02}
	A=\frac{\partial n_0(\left \langle T \right \rangle,\left \langle S \right \rangle,\lambda)}{\partial \left \langle T \right \rangle},B=\frac{\partial n_0(\left \langle T \right \rangle,\left \langle S \right \rangle,\lambda)}{\partial \left \langle S \right \rangle} 
\end{equation}
where the average refractive index $n_0(\left \langle T \right \rangle,\left \langle S \right \rangle,\lambda)$ can be calculated as \cite{ref18}   
\begin{equation} \label{equ:03}
	\begin{split}
	&n_0(\left \langle T \right \rangle,\left \langle S \right \rangle,\lambda)=a_0+(a_1+a_2\left \langle T \right \rangle+a_3\left \langle T \right \rangle ^2)\left \langle S \right \rangle\\
	&+a_4\left \langle T \right \rangle ^2+(a_5+a_6\left \langle S \right \rangle+a_7\left \langle T \right \rangle)\lambda^{-1}+a_8\lambda^{-2}+a_9\lambda^{-3}
	\end{split}
\end{equation}
where $a_0=1.31405$, $a_1=1.779\times 10^{-4}$, $a_2=-1.05\times 10^{-6}$, $a_3=1.6\times 10^{-8}$, $a_4=-2.02\times 10^{-6}$, $a_5=15.868$, $a_6=0.01155$, $a_7=-0.00423$, $a_8=-4382$ and $a_9=1.1455\times 10^{6}$.\\
\indent Each component in Eq. (\ref{equ:01}) is given by
\begin{equation} \label{equ:04}
	\begin{split}
	 \Phi_i(\kappa)&=\frac{\beta_0}{4\pi}\varepsilon^{-1/3}\kappa^{-11/3}\chi_i[1+\Delta_{1,i}\kappa^{0.61}-\Delta_{2,i}\kappa^{0.55}]\\
  &\times \exp(-\Delta_{3,i}\kappa^2),i\in\left\{T,S,TS\right\}	
	\end{split}
\end{equation}
where $\beta_0=0.72$ is the Obukhov-Corrsin constant and $\varepsilon$ is the energy dissipation rate that ranges from $10^{-10}$ to $10^{-1} \mathrm{m^2 s^{-3}}$. $\chi_T$ is the average temperature dissipation rate and ranges from $10^{-10}$ to $10^{-4} \mathrm{K^2 s^{-1}}$, $\chi_S=d_r H^{-2} \chi_T$ and $\chi_{TS}=0.5(1+d_r) H^{-1} \chi_T$. $d_r$ is the eddy diffusivity ratio, and $H$ is the temperature-salinity gradient ratio. Detailed description of these parameters is shown in \cite{ref18}.\\
\indent For the sake of mathematical simplicity, some notations are indicated as $\Delta_{1,i}=21.61\eta^{0.61}c_i^{0.02}$, $\Delta_{2,i}=18.18\eta^{0.55}c_i^{0.04}$, and $\Delta_{3,i}=174.90\eta^{2}c_i^{0.96}$, where $\eta$ is the Kolmogorov microscale (subsequently simplify to “microscale”), $c_i$ are the non-dimensional parameters defined as
\begin{equation}
	\label{equ:05}
\begin{cases}
\begin{split}
c_{T}&=0.072^{4 / 3} \beta_{0} \mathrm{Pr}^{-1} \\
c_{S}&=0.072^{4 / 3} \beta_{0} \mathrm{Sc}^{-1} \\
c_{TS}&=0.072^{4 / 3} \beta_{0}(\mathrm{Pr}+\mathrm{Sc}) \mathrm{Pr}^{-1} \mathrm{Sc}^{-1} 
\end{split}
\end{cases}
\end{equation}
where $\mathrm{Pr}$ is the Prandtl number, $\mathrm{Sc}$ is the Schmidt number.
\subsection{On-axis Scintillation Index Derivation}
In the context of Rytov theory, the SI $\sigma_I^2$ is approximately equal to the log-irradiance variance $\sigma_{\ln_{}{I}}^2$ in the weak fluctuation regime, i.e., $\sigma_I^2 \cong \sigma_{\ln_{}{I}}^2 = 4 \sigma_{\chi}^2$. The log-amplitude variance $\sigma_{\chi}^2$ is calculated by high-order statistical quantities of optical field $U(r,L)$. For a Gaussian beam with finite beam width, it exhibits both on-axis SI and off-axis SI. In the case of strict transceiver alignment, on-axis SI dominates the irradiance fluctuations. For homogeneous and isotropic turbulence, the on-axis SI takes the form \cite{ref04}
\begin{equation}
	\label{equ:06}
 \begin{split}
	\sigma_G^2(0,L) &=8\pi^2 k^2 L \\
 &\times \int_{0}^{1}\int_{0}^{\infty}\kappa\Phi_n(\kappa)\left[1-\cos\left(\frac{L\kappa^2}{k}\xi(1-\bar{\Theta}\xi)\right)\right] \\
 &\times \exp\left(-\frac{\Lambda L \kappa^2 \xi^2}{k}\right)d\kappa d\xi
 \end{split}
\end{equation}
where $\xi = 1- z/L$. $k = 2\pi/\lambda$ is the wave number and $L$ is the link distance and $\bar{\Theta} = 1-\Theta$. $\Theta =\Theta_0 / (\Theta_0^2+\Lambda_0^2)$ and $\Lambda=\Lambda_0 /(\Theta_0^2+\Lambda_0^2)$ are transmitter beam parameters of Gaussian beam, while $\Theta_0 = 1-L/F_0$ and $\Lambda_0 = 2L /(k W_0^2)$ are receiver beam parameters, where $W_0$ is the beam radius and $F_0$ is the radius of curvature of the phase front at the transmitter.\\
\indent Using Euler’s formula, writing the cosine function in Eq. (\ref{equ:06}) as $\cos{x}=\mathrm{Re}\left(e^{-jx}\right)$  and substituting Eq. (\ref{equ:01}) and Eq. (\ref{equ:04}) into Eq. (\ref{equ:06}) lead to
\begin{equation} \label{equ:07}
\begin{split}
	\sigma_G^2(0,L) &=2\pi L \beta_0 k^2 \varepsilon^{-1/3} \\
 & \times \left[A^2\chi_T K_T +B^2\chi_S K_S +2AB\chi_{TS} K_{TS}\right]
 \end{split}
\end{equation}
\begin{figure*}[!t]
    \begin{equation}
        \label{equ:08}
	\begin{split}
		K_i &=\int_{0}^{1}\int_{0}^{\infty}\kappa^{-8/3}\left[1+\frac{4.47}{c_i^{0.2728}}\left(\frac{\kappa}{\kappa_{\eta,i}}\right)^{0.61}-\frac{4.39}{c_i^{0.224}}\left(\frac{\kappa}{\kappa_{\eta,i}}\right)^{0.55}\right]\exp\left\{-(1+\Lambda \xi^2 Q_{\eta,i})\left(\frac{\kappa}{\kappa_{\eta,i}}\right)^2\right\}d\kappa d\xi \\
& -\mathrm{Re}\int_{0}^{1}\int_{0}^{\infty}\kappa^{-8/3}\left[1+\frac{4.47}{c_i^{0.2728}}\left(\frac{\kappa}{\kappa_{\eta,i}}\right)^{0.61}-\frac{4.39}{c_i^{0.224}}\left(\frac{\kappa}{\kappa_{\eta,i}}\right)^{0.55}\right]\exp\left\{-\left(\frac{\kappa}{\kappa_{\eta,i}}\right)^2\left[(1+\Lambda \xi^2 Q_{\eta,i})+jQ_{\eta,i}\xi(1-\bar{\Theta}\xi)\right]\right\}d\kappa d\xi.
	\end{split}
\end{equation} 
\hrulefill
\vspace*{4pt}
\end{figure*}
\begin{figure*}[!t]
\begin{equation} \label{equ:09}
\begin{split}
 K_i=\kappa_{\eta, i}^{-5 / 3}\left\{\mathrm {Re}\left[\begin{array}{l}
3.34 \int_{0}^{1}\left[1+\Lambda \xi^{2} Q_{\eta, i}+j Q_{\eta, i} \xi(1-\bar{\Theta} \xi)\right]^{\frac{5}{6}} d \xi \\
+\frac{7.95}{c_{i}^{0.2728}} \int_{0}^{1}\left[1+\Lambda \xi^{2} Q_{\eta, i}+j Q_{\eta, i} \xi(1-\bar{\Theta} \xi)\right]^{\frac{5}{6}-\frac{0.61}{2}} d \xi \\
-\frac{7.89}{c_{i}^{0.224}} \int_{0}^{1}\left[1+\Lambda \xi^{2} Q_{\eta, i}+j Q_{\eta, i} \xi(1-\bar{\Theta} \xi)\right]^{\frac{5}{6}-\frac{0.55}{2}} d \xi
\end{array}\right]-\left[\begin{array}{l}
3.34 \int_{0}^{1}\left(1+\Lambda \xi^{2} Q_{\eta, i}\right)^{\frac{5}{6}} d \xi \\
+\frac{7.95}{c_{i}^{0.2728}} \int_{0}^{1}\left(1+\Lambda \xi^{2} Q_{\eta, i}\right)^{\frac{5}{6}-\frac{0.61}{2}} d \xi \\
-\frac{7.89}{c_{i}^{0.224}} \int_{0}^{1}\left(1+\Lambda \xi^{2} Q_{\eta, i}\right)^{\frac{5}{6}-\frac{0.55}{2}} d \xi
\end{array}\right]\right\}
\end{split}
\end{equation}
\hrulefill
\vspace*{4pt}
\end{figure*}
where $K_i, i\in\left\{T,S,TS\right\}$ has the identical form as Eq. (\ref{equ:08}).\\
\indent Here $Q_{\eta,i}=L\kappa_{\eta,i}^2/k,i\in \left\{T,S,TS\right\}$ is the non-dimensional parameter and $\kappa_{\eta,i} = \Delta_{3,i}^{-1/2}$ is the corresponding spatial frequency. \\
\indent Next, by using the integral equation $\int_{0}^{\infty}e^{-st}t^{x-1}dt=\Gamma(x)/s^x$ to handle the inside integration for variable $\kappa$, Eq. (\ref{equ:08}) can be expressed as Eq. (\ref{equ:09}).\\
\indent There are two kinds integrals need to be solved in Eq. (\ref{equ:09}), we write them as 
\begin{equation}
	\label{equ:10}
\begin{cases}
\begin{split}
I_1 &= \int_{0}^{1}\left[1+\Lambda \xi^{2} Q_{\eta, i}+j Q_{\eta, i} \xi(1-\bar{\Theta} \xi)\right]^{\frac{5}{6}} d \xi \\
I_2 &= \int_{0}^{1}\left(1+\Lambda \xi^{2} Q_{\eta, i}\right)^{\frac{5}{6}} d \xi = { }_{2} F_{1}(-\frac{5}{6},\frac{1}{2};\frac{3}{2};-\Lambda Q_{\eta,i})
\end{split}
\end{cases}
\end{equation}
\indent $I_2$ can be directly simplified by the hypergeometric function ${ }_{2} F_{1}(a,b;c;x)$ (Eq. (Appendix I.H3) in \cite{ref04}). As for $I_1$, we first use the Taylor series expansion $(1+x)^a = \sum_{n=0}^{\infty} \frac{1}{n!}\frac{a!}{(a-n)!} x^n$ to write the integrand as a series sum. Next, we express the coefficient before integral as the Pochhammer symbol $(a)_n$ by using Eq. (Appendix I.5) in \cite{ref04}. Then, similar to $I_2$, we turn the integral into the form of ${ }_{2} F_{1}(a,b;c;x)$ as
\begin{equation}
	\label{equ:11}
 \begin{split}
 I_1 &= \int_{0}^{1}\left[1+\Lambda \xi^{2} Q_{\eta, i}+j Q_{\eta, i} \xi(1-\bar{\Theta} \xi)\right]^{\frac{5}{6}} d \xi \\
 &=  \sum_{n=0}^{\infty} \frac{(-5/6)_n}{n!}(-j Q_{\eta, i})^n \int_{0}^{1}\xi^n \left[(1-(\bar{\Theta}+j \Lambda)\xi \right]^n d \xi \\
 &= \sum_{n=0}^{\infty} \frac{(-5/6)_n (1)_n}{(2)_n n!}(-j Q_{\eta, i})^n \times { }_{2} F_{1}(-n,n+1;n+2;\bar{\Theta}+j \Lambda)
 \end{split}
\end{equation}
\indent After solving all integrals in Eq. (\ref{equ:09}) with the same procedures, we finally write Eq. (\ref{equ:09}) in the form of Eq. (\ref{equ:12}), where $F(n) = { }_{2} F_{1}(-n,n+1;n+2;\bar{\Theta}+j \Lambda)$.\\
\begin{figure*}[!t]
\begin{equation} \label{equ:12}
\begin{split}
K_i=\kappa_{\eta, i}^{-5/3} \left\{ \mathrm{Re} \left[\begin{array}{l} 
3.34 \sum_{n=0}^{\infty} \frac{(-\frac{5}{6})_{n} (1)_{n}}{(2)_{n} n !} (-j Q_{\eta, i})^{n} \times F(n)  \\ 
+\frac{7.95}{c_{i}^{0.2728}} \sum_{n=0}^{\infty}\frac{(-\frac{5}{6} + \frac{0.61}{2})_{n} (1)_{n}}{(2)_{n} n !} (-j Q_{\eta, i})^{n} \times F(n) \\ 
-\frac{7.89}{c_{i}^{0.224}} \sum_{n=0}^{\infty}\frac{(-\frac{5}{6} +\frac{0.55}{2})_{n} (1)_{n}}{(2)_{n} n!}(-j Q_{\eta, i})^{n} \times F(n) 
\end{array}\right]-\left[\begin{array}{l} 
3.34 { }_{2} F_{1}(-\frac{5}{6},\frac{1}{2};\frac{3}{2};-\Lambda Q_{\eta,i}) \\ 
+\frac{7.95}{c_{i}^{0.2728}} { }_{2} F_{1}(-\frac{5}{6} + \frac{0.61}{2},\frac{1}{2};\frac{3}{2};-\Lambda Q_{\eta,i}) \\ 
-\frac{7.89}{c_{i}^{0.224}} { }_{2} F_{1}(-\frac{5}{6} + \frac{0.55}{2},\frac{1}{2};\frac{3}{2};-\Lambda Q_{\eta,i}) 
\end{array} \right] \right\} , Q_{\eta,i}<1
\end{split}
\end{equation}
\hrulefill
\vspace*{4pt}
\end{figure*}
\indent The restriction $Q_{\eta, i}<1$ in Eq. (\ref{equ:12}) can be extended to be valid for all $Q_{\eta, i}$, where the following approximation is used \cite{ref24}
\begin{equation} \label{equ:13}
\begin{split}
_2F_1(-n,n+1;n+2;x) \approx \left(1-\frac{2}{3}x \right)^n, \left| x \right| <1  
\end{split}
\end{equation}
\indent Then sum the series in Eq. (\ref{equ:12}) to obtain Eq. (\ref{equ:14}).\\
\begin{figure*}[!t]
\begin{equation} \label{equ:14}
\begin{split}
K_i=\kappa_{\eta, i}^{-5 / 3}\left\{ \mathrm {Re}\left[\begin{array}{l}
3.34 { }_{2} F_{1} \left( -\frac{5}{6},1;2;-j Q_{\eta,i}\left[ 1-\frac{2}{3}(\bar{\Theta}+j \Lambda)\right]  \right)  \\
+\frac{7.95}{c_{i}^{0.2728}} { }_{2} F_{1} \left( -\frac{5}{6}+ \frac{0.61}{2},1;2;-j Q_{\eta,i}\left[ 1-\frac{2}{3}(\bar{\Theta}+j \Lambda)\right]  \right)  \\
-\frac{7.89}{c_{i}^{0.224}} { }_{2} F_{1} \left( -\frac{5}{6}+ \frac{0.55}{2},1;2;-j Q_{\eta,i}\left[ 1-\frac{2}{3}(\bar{\Theta}+j \Lambda)\right]  \right) 
\end{array}\right] -\left[\begin{array}{l}
3.34 { }_{2} F_{1}(-\frac{5}{6},\frac{1}{2};\frac{3}{2};-\Lambda Q_{\eta,i}) \\
+\frac{7.95}{c_{i}^{0.2728}} { }_{2} F_{1}(-\frac{5}{6} + \frac{0.61}{2},\frac{1}{2};\frac{3}{2};-\Lambda Q_{\eta,i}) \\
-\frac{7.89}{c_{i}^{0.224}} { }_{2} F_{1}(-\frac{5}{6} + \frac{0.55}{2},\frac{1}{2};\frac{3}{2};-\Lambda Q_{\eta,i})
\end{array}\right] \right\} 
\end{split}
\end{equation}
\hrulefill
\vspace*{4pt}
\end{figure*}
\indent To produce a simpler expression, we have used the identity \cite{ref25}
\begin{equation}
	\label{equ:15}
	\begin{split}
		_2F_1 (1-a,1;2;-x) = \frac{(1+x)^a -1}{ax}
	\end{split}
\end{equation}
\indent Substituting Eq. (\ref{equ:15}) into Eq. (\ref{equ:14}) yields Eq. (\ref{equ:16}).\\
\begin{figure*}[!t]
\begin{equation} \label{equ:16}
\begin{split}
K_i=\kappa_{\eta, i}^{-5 / 3}\left \{ \mathrm {Re}\left[\begin{array}{l}
5.47 \frac{\left(1+\frac{Q_{\eta,i}}{3}\left[2\Lambda+j(1+2\Theta)\right]\right)^{\frac{11}{6}}-1}{Q_{\eta,i}\left[ 2\Lambda+j(1+2\Theta)\right]}  \\
+\frac{15.60}{c_{i}^{0.2728}} \frac{\left(1+\frac{Q_{\eta,i}}{3}\left[2\Lambda+j(1+2\Theta)\right]\right)^{\frac{11}{6} \frac{0.61}{2} }-1}{Q_{\eta,i}\left[ 2\Lambda+j(1+2\Theta)\right]}  \\
-\frac{15.19}{c_{i}^{0.224}} \frac{\left(1+\frac{Q_{\eta,i}}{3}\left[2\Lambda+j(1+2\Theta)\right]\right)^{\frac{11}{6} -\frac{0.55}{2} }-1}{Q_{\eta,i}\left[ 2\Lambda+j(1+2\Theta)\right ]} 
\end{array}\right] -\left[\begin{array}{l}
3.34 { }_{2} F_{1}(-\frac{5}{6},\frac{1}{2};\frac{3}{2};-\Lambda Q_{\eta,i}) \\
+\frac{7.95}{c_{i}^{0.2728}} { }_{2} F_{1}(-\frac{5}{6} + \frac{0.61}{2},\frac{1}{2};\frac{3}{2};-\Lambda Q_{\eta,i}) \\
-\frac{7.89}{c_{i}^{0.224}} { }_{2} F_{1}(-\frac{5}{6} + \frac{0.55}{2},\frac{1}{2};\frac{3}{2};-\Lambda Q_{\eta,i})
\end{array}\right] \right\} 
\end{split}
\end{equation}
\hrulefill
\vspace*{4pt}
\end{figure*}
\indent In view of the relation $\kappa_{\eta,i}^{-5/3} = \left( \frac{L}{k} \right)^{\frac{5}{6}} Q_{\eta,i}^{-\frac{5}{6}}$, Eq. (\ref{equ:07}) can be equivalent to
\begin{equation} \label{equ:17}
\begin{split}
\sigma_G^2(0,L) &\approx 2\pi L^{11/6} \beta_0 k^{7/6} \varepsilon^{-1/3} \\
 & \times \left[A^2\chi_T M_T +B^2\chi_S M_S +2AB\chi_{TS} M_{TS}\right]
\end{split}
\end{equation}
where $M_i, i\in\left\{T,S,TS\right\}$ has the identical form as Eq. (\ref{equ:18}).\\
\begin{figure*}[!t]
\begin{equation} \label{equ:18}
\begin{split}
M_i=Q_{\eta, i}^{-5/6}\left \{ \mathrm {Re}\left[\begin{array}{l}
5.47 \frac{\left(1+\frac{Q_{\eta,i}}{3}\left[2\Lambda+j(1+2\Theta)\right]\right)^{\frac{11}{6}}-1}{Q_{\eta,i}\left[ 2\Lambda+j(1+2\Theta)\right]}  \\
+\frac{15.60}{c_{i}^{0.2728}} \frac{\left(1+\frac{Q_{\eta,i}}{3}\left[2\Lambda+j(1+2\Theta)\right]\right)^{\frac{11}{6} \frac{0.61}{2} }-1}{Q_{\eta,i}\left[ 2\Lambda+j(1+2\Theta)\right]}  \\
-\frac{15.19}{c_{i}^{0.224}} \frac{\left(1+\frac{Q_{\eta,i}}{3}\left[2\Lambda+j(1+2\Theta)\right]\right)^{\frac{11}{6} -\frac{0.55}{2} }-1}{Q_{\eta,i}\left[ 2\Lambda+j(1+2\Theta)\right]} 
\end{array}\right] -\left [\begin{array}{l}
3.34 { }_{2} F_{1}(-\frac{5}{6},\frac{1}{2};\frac{3}{2};-\Lambda Q_{\eta,i}) \\
+\frac{7.95}{c_{i}^{0.2728}} { }_{2} F_{1}(-\frac{5}{6} + \frac{0.61}{2},\frac{1}{2};\frac{3}{2};-\Lambda Q_{\eta,i}) \\
-\frac{7.89}{c_{i}^{0.224}} { }_{2} F_{1}(-\frac{5}{6} + \frac{0.55}{2},\frac{1}{2};\frac{3}{2};-\Lambda Q_{\eta,i})
\end{array} \right ] \right \} 
\end{split}
\end{equation}
\hrulefill
\vspace*{4pt}
\end{figure*}
\indent For the sake of form simplicity, we define a coefficient $\sigma_1^2 = 2.96 \beta_0 \varepsilon^{-1/3} k^{7/6} L^{11/6}$, thus Eq. (\ref{equ:17}) can be simplified as
\begin{equation}
	\label{equ:19}
	\begin{split}
		\sigma_G^2(0,L) &\approx 2.12\sigma_1^2
  \left[A^2\chi_T M_T +B^2\chi_S M_S +2AB\chi_{TS} M_{TS}\right]
	\end{split}
\end{equation}
\indent The limiting case of a plane wave is obtained by specifying $\Theta = 1$ and $\Lambda =0$, which is expressed as 
\begin{equation}
	\label{equ:20}
	\begin{split}
		&\sigma_{PL}^2(L) \approx 2.12\sigma_1^2 \\
  &\times \left[A^2\chi_T M_{T,pl} +B^2\chi_S M_{S,pl} +2AB\chi_{TS} M_{TS,pl}\right]
	\end{split}
\end{equation}
where 
\begin{equation} \label{equ:21}
\begin{split}
M_{i,pl} &= \mathrm {Re}\left[\begin{array}{l}
5.47 \frac{(1+jQ_{\eta,i})^{\frac{11}{6}}-1}{Q_{\eta,i}^{\frac{11}{6}}[j3]}  \\
+\frac{15.60}{c_{i}^{0.2728}} \frac{(1+jQ_{\eta,i})^{\frac{11}{6}-\frac{0.61}{2} }-1}{Q_{\eta,i}^{\frac{11}{6}}[j3]}  \\
-\frac{15.19}{c_{i}^{0.224}} \frac{(1+jQ_{\eta,i})^{\frac{11}{6}-\frac{0.55}{2} }-1}{Q_{\eta,i}^{\frac{11}{6}}[j3]} 
\end{array}\right] \\
&-Q_{\eta, i}^{-5/6} \left[3.34 +\frac{7.95}{c_{i}^{0.2728}} -\frac{7.89}{c_{i}^{0.224}} \right] 
\end{split}
\end{equation}
\indent Analogously, the spherical wave limit is obtained by specifying $\Theta = \Lambda =0$, which is given by 
\begin{equation}
	\label{equ:22}
	\begin{split}
		&\sigma_{SP}^2(L) \approx 2.12\sigma_1^2 \\
  &\times \left[A^2\chi_T M_{T,sp} +B^2\chi_S M_{S,sp} +2AB\chi_{TS} M_{TS,sp}\right]
	\end{split}
\end{equation}
where 
\begin{equation} \label{equ:23}
\begin{split}
M_{i,sp} &= \mathrm {Re}\left[\begin{array}{l}
5.47 \frac{(1+jQ_{\eta,i}/3)^{\frac{11}{6}}-1}{jQ_{\eta,i}^{\frac{11}{6}}}  \\
+\frac{15.60}{c_{i}^{0.2728}} \frac{(1+jQ_{\eta,i}/3)^{\frac{11}{6}-\frac{0.61}{2} }-1}{jQ_{\eta,i}^{\frac{11}{6}}}  \\
-\frac{15.19}{c_{i}^{0.224}} \frac{(1+jQ_{\eta,i}/3)^{\frac{11}{6}-\frac{0.55}{2} }-1}{jQ_{\eta,i}^{\frac{11}{6}}} 
\end{array}\right] \\
&-Q_{\eta, i}^{-5/6} \left[3.34 +\frac{7.95}{c_{i}^{0.2728}} -\frac{7.89}{c_{i}^{0.224}} \right] 
\end{split}
\end{equation}
\indent Specially, if we allow the microscale vanish (i.e., let $\eta \to 0$ $(Q_{\eta, i} \to \infty)$ ), it is obvious to see that Eq. (\ref{equ:19}) reduce to its Rytov variance
\begin{equation} \label{equ:24}
\begin{split}
\sigma_{B}^2(L) &= 2.12\sigma_1^2
  (A^2\chi_T +B^2\chi_S  +2AB\chi_{TS} ) \\
&\times \left[0.73 \mathrm{Re} \left( [2\Lambda +j(1+2 \Theta)]^{5/6} \right) -1.25\Lambda^{5/6} \right] 
\end{split}
\end{equation}
\indent Based on Eq. (\ref{equ:24}), the limiting results of plane and spherical waves can be obtained by specifying $\Theta = 1,\Lambda =0$, and $\Theta = \Lambda =0$, respectively, which are coincident with Eq. (19) and Eq. (31) in \cite{ref19}. \\
\begin{figure}[ht!]
    \centering
    \includegraphics[width=0.48\textwidth]{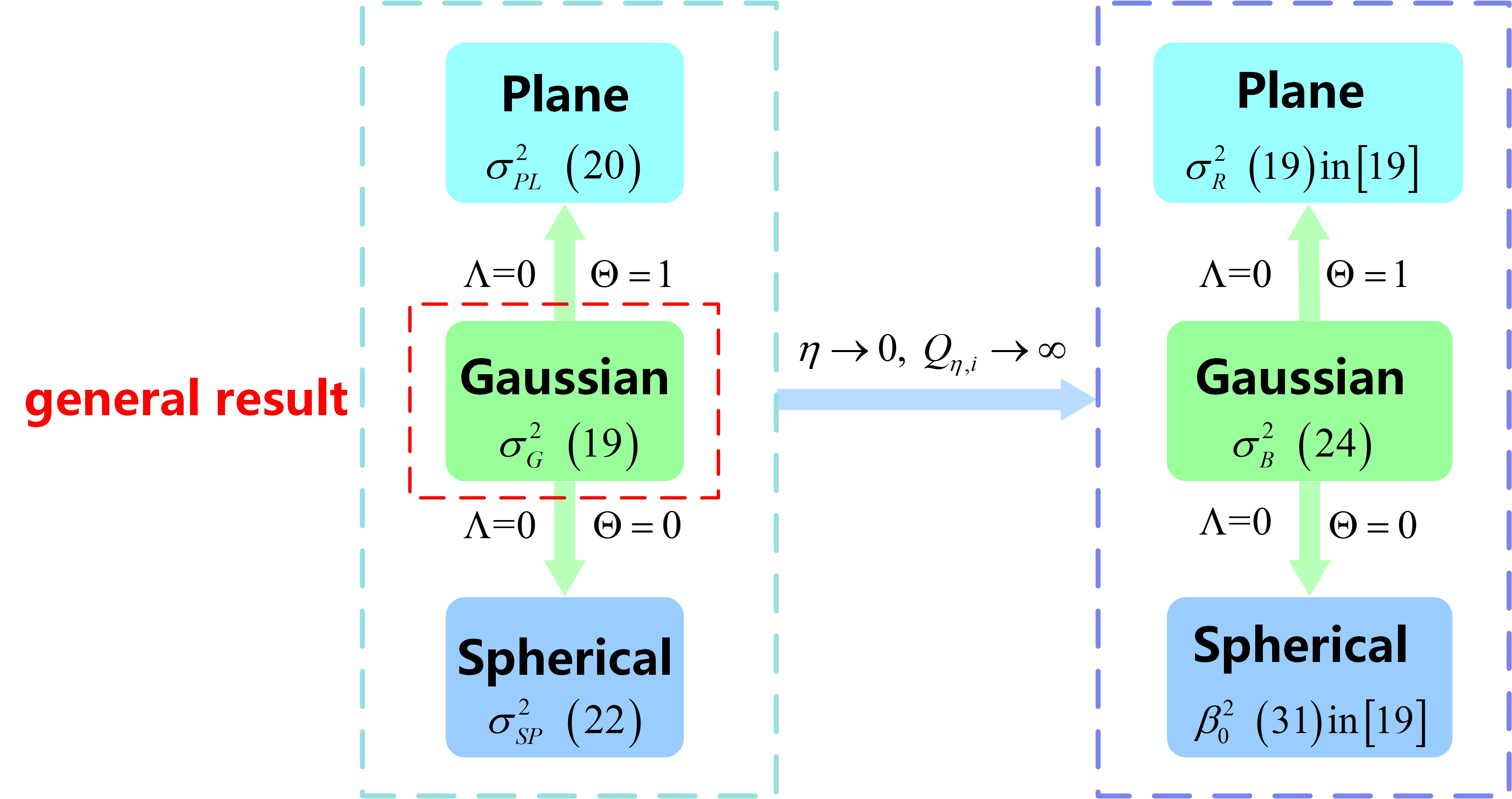}
    \caption{The SI of Gaussian beam, the comparison of this work and previous works.}
    \label{fig:01}
\end{figure}
\indent The relation between our work and previous works is shown in Figure~\ref{fig:01}. It is obvious that Eq. (\ref{equ:19}) is a more general result and other five results can be indirectly obtained from it through specific values.
\section{Bit Error Rate Performance Analysis of UWOC} \label{sec:BER}
Under weak irradiance fluctuations, the log-normal probability density function (PDF) of the random fading of optical irradiance intensity follows directly from the first Rytov approximation, which is expressed as \cite{ref04}
\begin{equation}
	\label{equ:25}
	\begin{split}
		f_I(I)=\frac{1}{I\sqrt{2\pi \sigma_I^2}}\exp \left [-\frac{(\ln I+0.5\sigma_I^2)^2}{2\sigma_I^2}\right],I>0 
	\end{split}
\end{equation}
where $\sigma_I^2$ represents the SI under weak oceanic turbulence. \\
\indent The mean signal-to-noise ratio (SNR) at the output of the detector in the case of a shot-noise limited system is written as \cite{ref04}
\begin{equation}
	\label{equ:26}
	\begin{split}
		\left \langle SNR \right \rangle =\frac{SNR_0}{\sqrt{1+\sigma_I^2(SNR_0)^2}}
	\end{split}
\end{equation}
where $SNR_0$ is the output SNR in the absence of optical turbulence. \\
\indent In the presence of oceanic turbulence, the error probability is considered as a conditional probability, which must be averaged over the PDF of random signals to determine the unconditional mean BER \cite{ref10}. For the UWOC system with the on-off-keying (OOK) modulation scheme, it is given by \cite{ref04}
\begin{equation}
	\label{equ:27}
	\begin{split}
		\left \langle BER \right \rangle =\frac{1}{2} \int_{0}^{\infty}f_I(I) \mathrm{erfc} \left(\frac{\left \langle SNR \right \rangle I}{2\sqrt{2}} \right)dI
	\end{split}
\end{equation}
where $\mathrm{erfc}(\cdot)$ is the complementary error function. In the absence of turbulence, Eq. (\ref{equ:27}) reduces to
\begin{equation}
	\label{equ:28}
	\begin{split}
		\left \langle BER \right \rangle =\frac{1}{2} \mathrm{erfc} \left(\frac{SNR_0}{2\sqrt{2}} \right)
	\end{split}
\end{equation}
\section{Numerical Results and Analysis} \label{sec:results}
This section presents the SI and average BER of UWOC system for various parameters in oceanic turbulence. Parameters are consistent with the following statements, unless otherwise stated in the figures' captions or on the plots. The collimated Gaussian beam with phase front radius of curvature $F_0 = \infty$ and the beam radius $W_0 = 3 \mathrm{mm}$ is chosen as the optical source, whose wavelength $\lambda = 532 \mathrm{nm}$. The link length is set as $L=20 \mathrm{m}$. For oceanic turbulence parameters, we select $\varepsilon = 10^{-2} \mathrm{m^2 s^{-3}}$ and $\chi_T = 10^{-5} \mathrm{K^2 s^{-1}}$ to conduct numerical analysis. The microscale $\eta$ of oceanic turbulence is in the range of $\eta = 6\times 10^{-5}\sim 10^{-2}\mathrm{m}$. Here we choose the typical value $\eta = 10^{-3} \mathrm{m}$. The average temperature on oceanic surface is nearly $\left \langle T \right \rangle = 20 ^{\circ} \mathrm{C}$ while the typical average salinity is related to sea area. Here we choose $\left \langle S \right \rangle = 34.9 \mathrm{ppt}$. The temperature-salinity gradient ratio $H$ is chosen as $-10^{\circ} \mathrm{C} \cdot \mathrm{ppt}^{-1}$.\\
\indent Numerical results are obtained under MATLAB environment. Firstly, we present the results obtained at the microscale $\eta \ne 0$, which exhibit significant differences from those obtained at the microscale $\eta \to 0$. This highlights the necessity of considering finite-microscale cases. Eq. (8) in \cite{ref21} is only applicable to specific limiting case $\eta \to 0$, i.e., Rytov variance results. But there is not such restriction in our work, indicating the universality of our expressions. It is worth noting that even both are Rytov variance results, approximate results in \cite{ref21} is not always consistent with our exact results. Detailed phenomena and analysis are shown in Section \ref{sec:4-1}. This demonstrates the accuracy of our derivation. Then, we study the SI under various beam, propagation and oceanic turbulence parameters. Specially, we take extra accounts for the correlation between SI and both different depths and oceans. Finally, on the basis of our precise results of SI and the log-normal PDF, we investigate the BER performance of UWOC systems.
\subsection{Numerical Results of the Scintillation Index} \label{sec:4-1}
\begin{figure}[t!]
	\centering
        \subfloat[]{
        \includegraphics[width=0.45\linewidth]{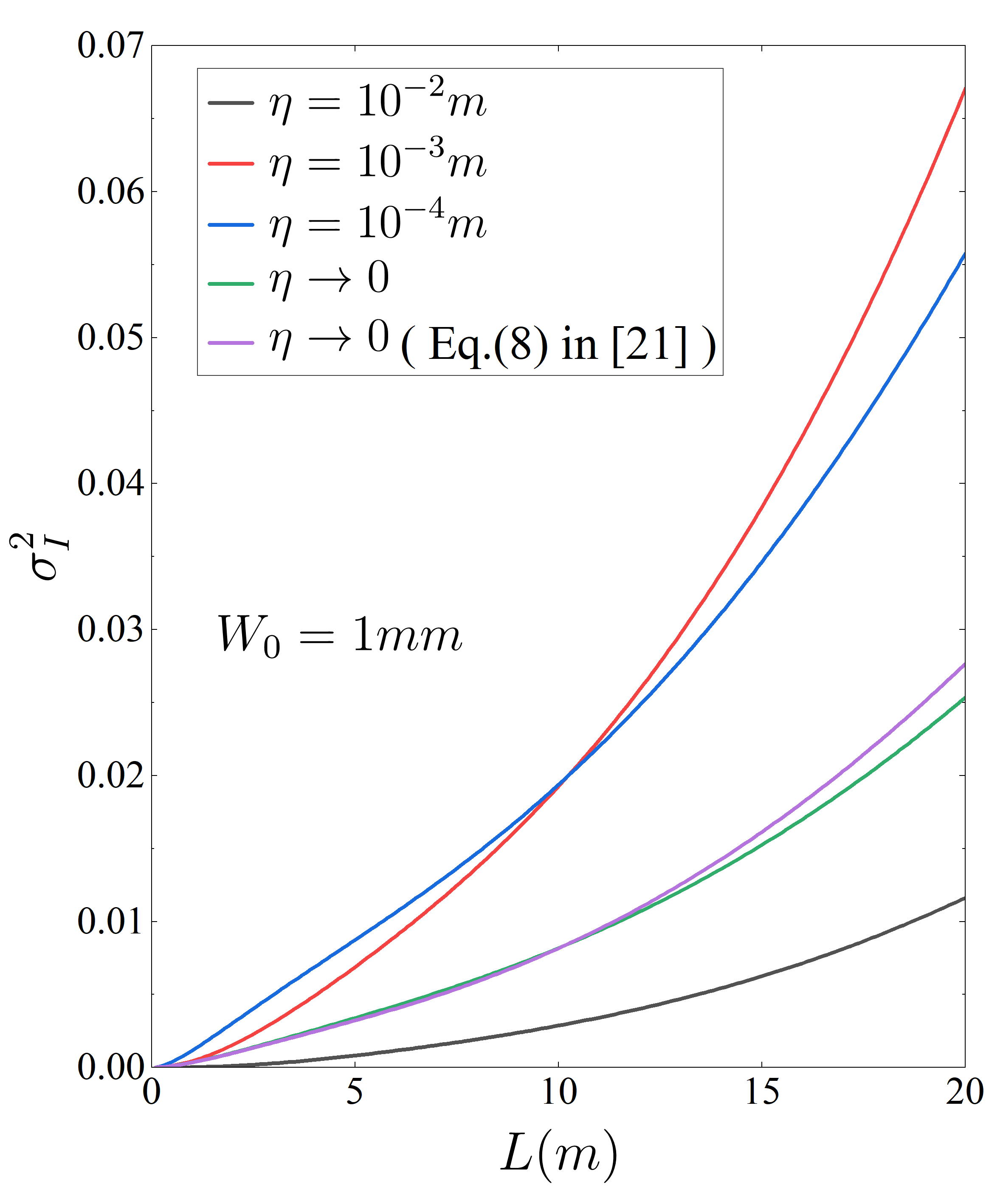} 
        \label{fig:02a} }
	\centering
        \subfloat[]{
        \includegraphics[width=0.45\linewidth]{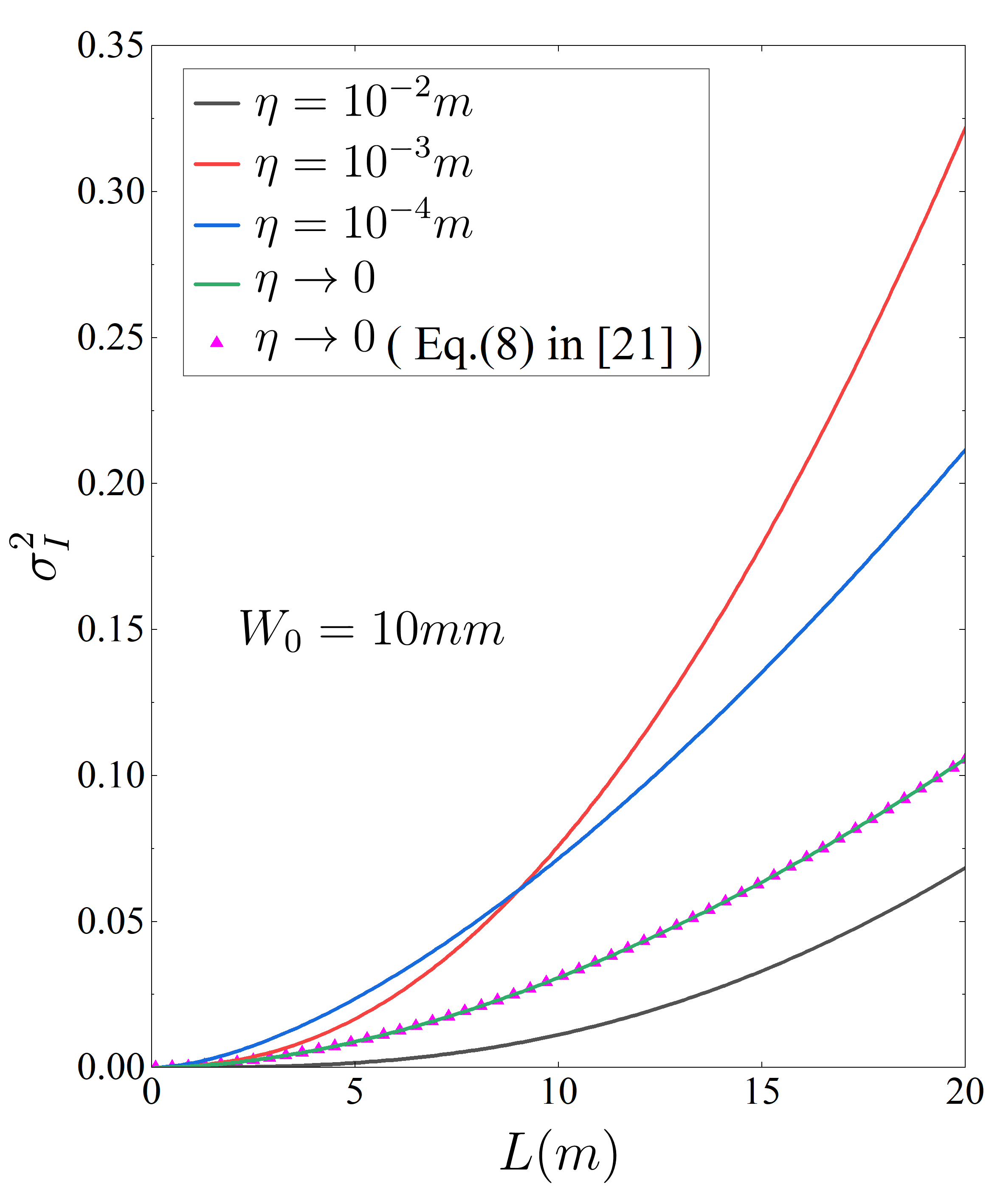} 
        \label{fig:02b} }
        \caption{The SI versus link length for different microscales in case (a) $W_0 = 1\mathrm{mm}$ and (b) $W_0 = 10\mathrm{mm}$.}
        \label{fig:02}
\end{figure}
\begin{figure}[h!]
	\centering
	\subfloat[]{
        \includegraphics[width=0.45\linewidth]{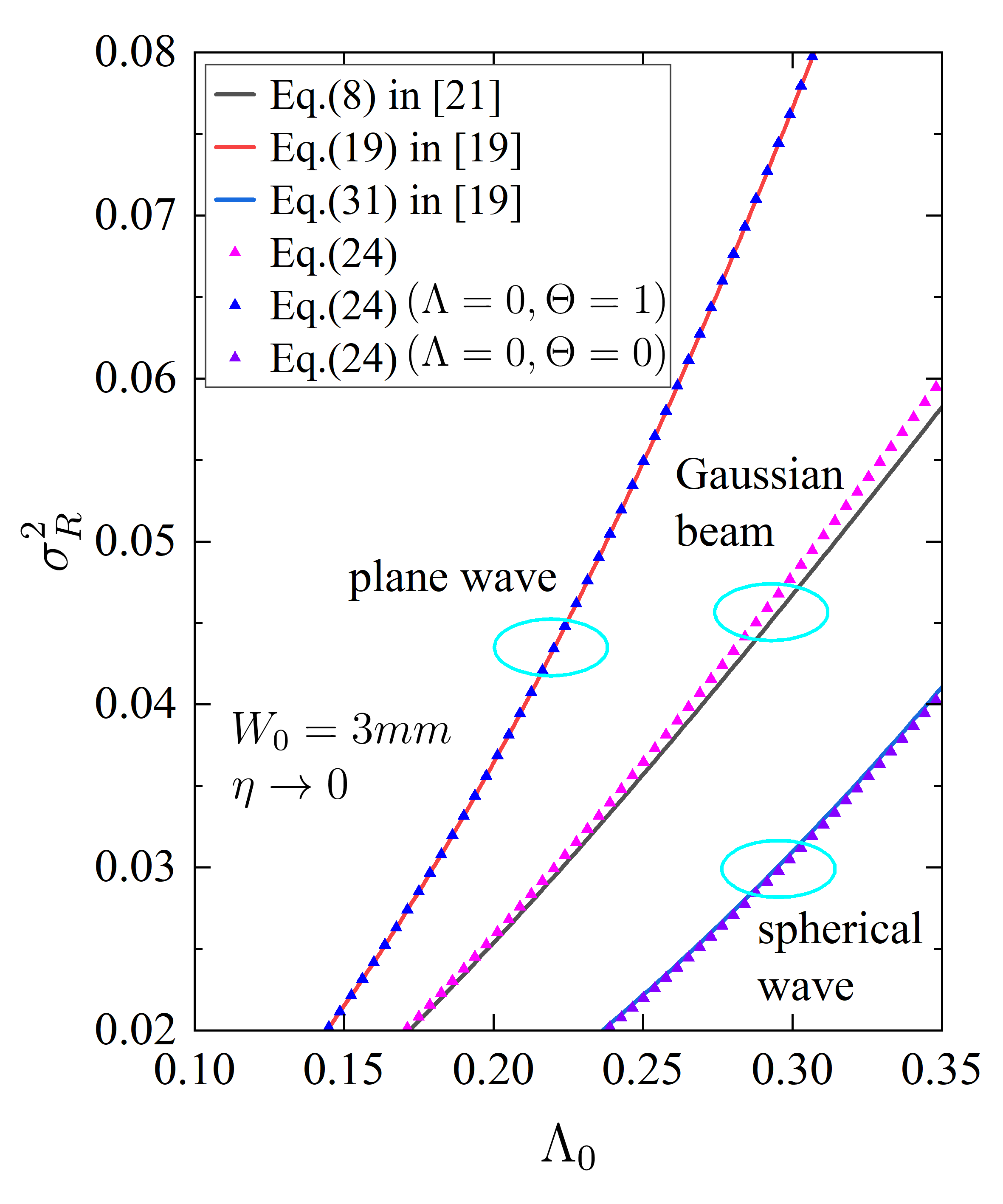} 
        \label{fig:03a} }
	\centering
        \subfloat[]{
        \includegraphics[width=0.45\linewidth]{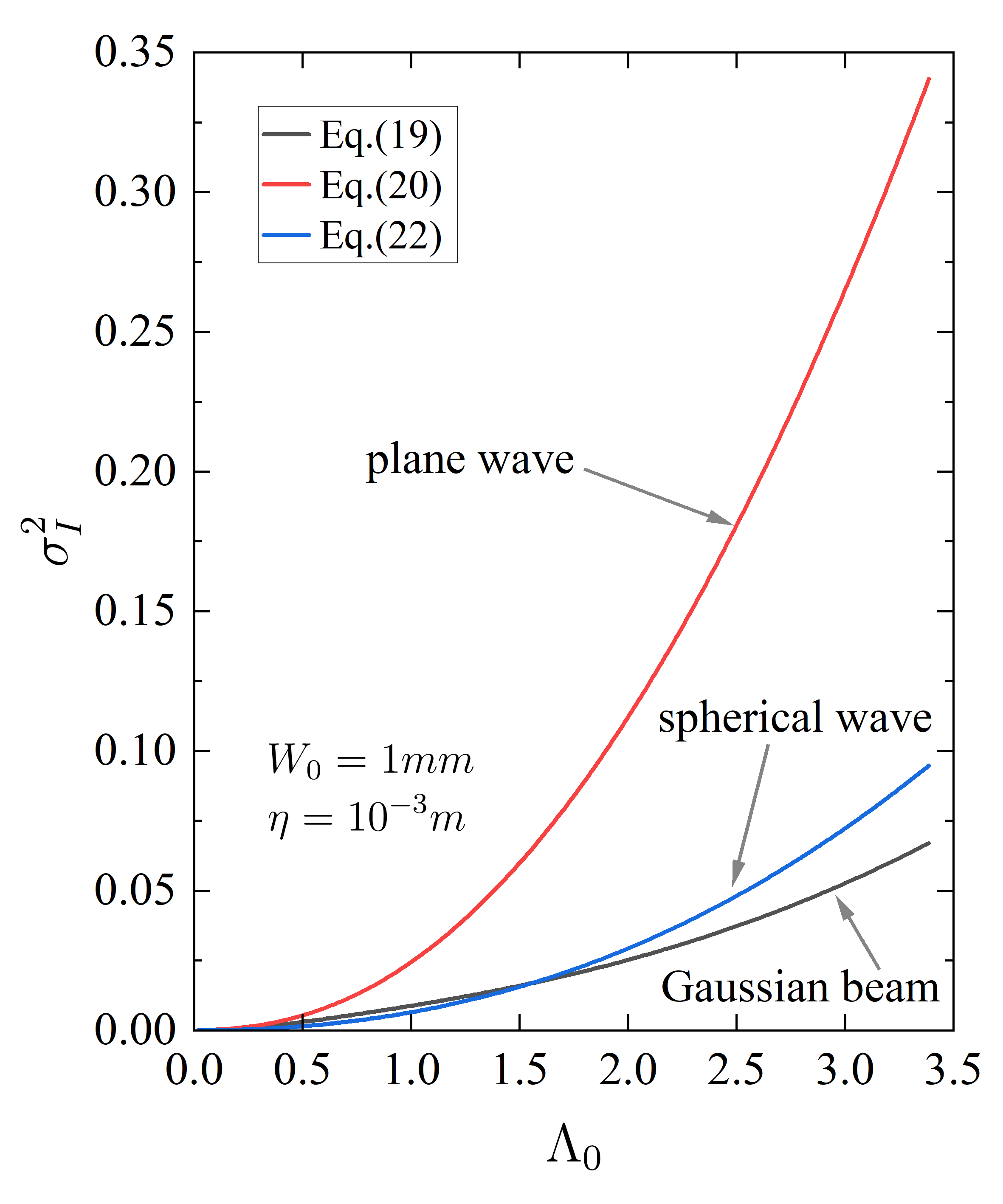} 
        \label{fig:03b} }
        \qquad
        \centering
	\subfloat[]{
        \includegraphics[width=0.45\linewidth]{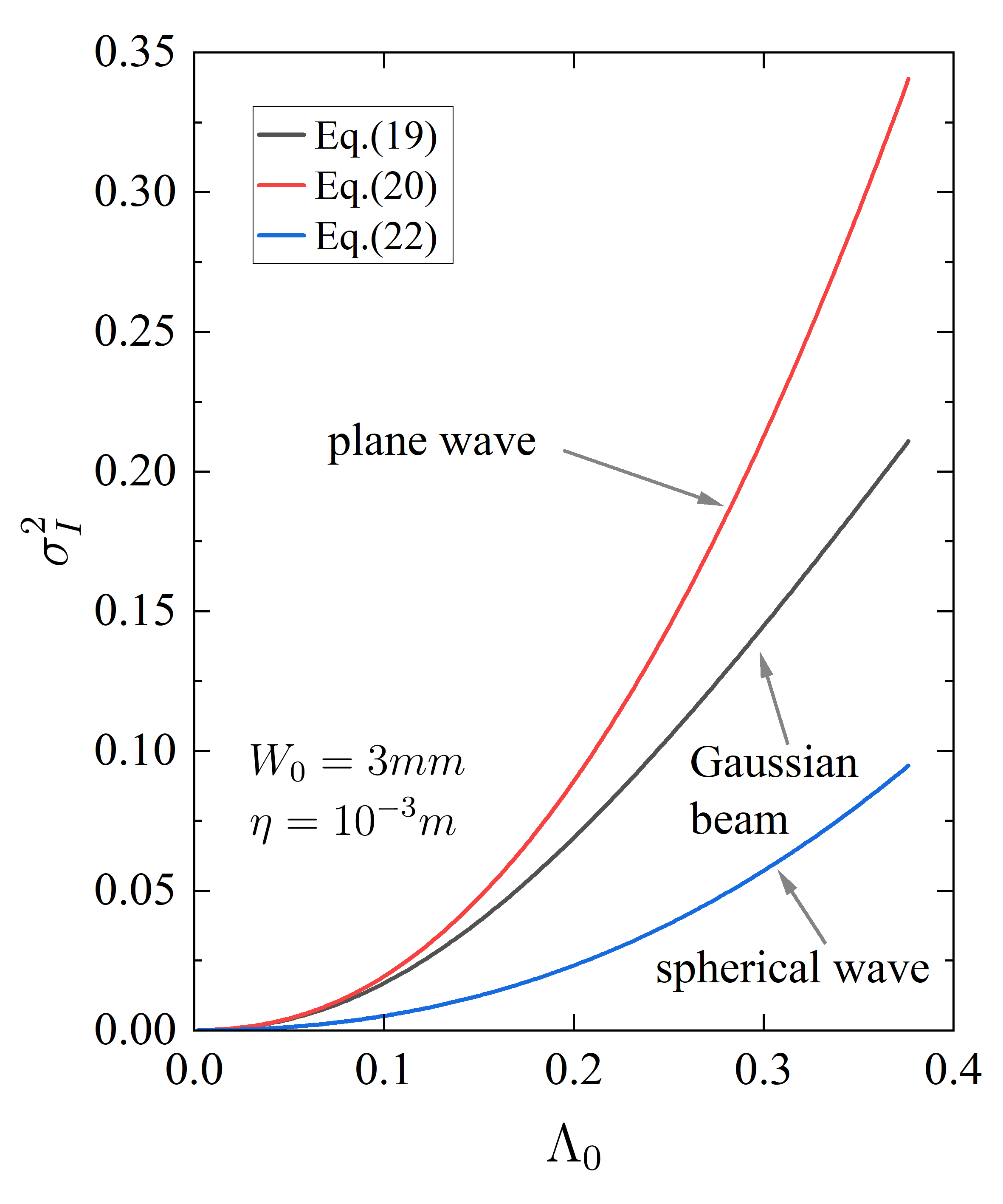} 
        \label{fig:03c} }
	\centering
        \subfloat[]{
        \includegraphics[width=0.45\linewidth]{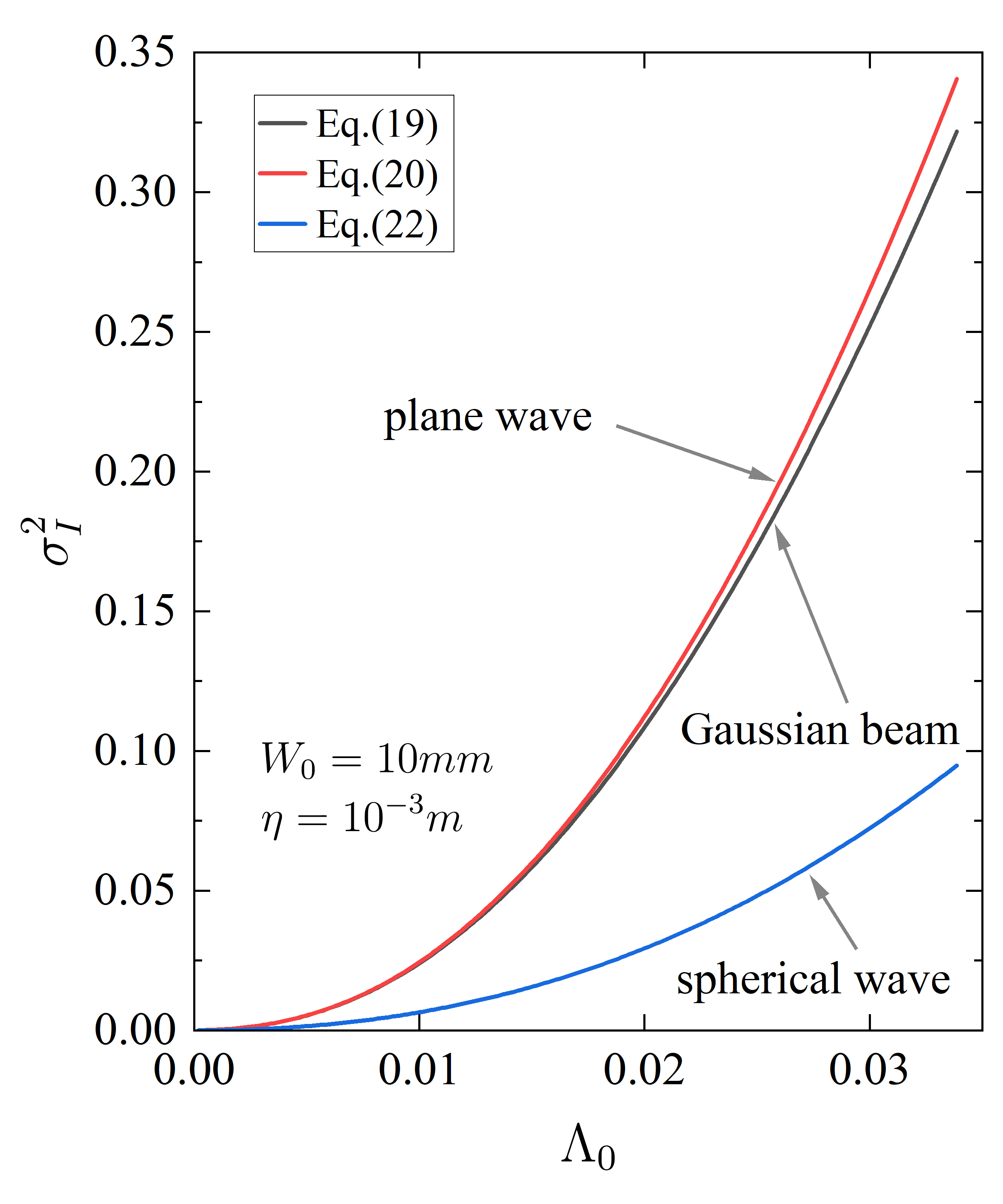} 
        \label{fig:03d} }
        \caption{The SI of three beams versus Fresnel ratio $\Lambda_0$ for different beam radius and microscales, (a) $W_0 = 3\mathrm{mm}, \eta \to 0$, (b) $W_0 = 1\mathrm{mm}, \eta = 10^{-3}\mathrm{m}$, (c) $W_0 = 3\mathrm{mm}, \eta = 10^{-3}\mathrm{m}$ and (d) $W_0 = 10\mathrm{mm}, \eta = 10^{-3}\mathrm{m}$.}
        \label{fig:03}
\end{figure}
\indent Figure~\ref{fig:02} depicts the variation of SI with link length, where the beam radius in Figure~\ref{fig:02a} is $W_0 =1 \mathrm{mm}$ and increases to $W_0 =10 \mathrm{mm}$ in Figure~\ref{fig:02b}. The black, red and blue curves correspond to the results of Eq. (\ref{equ:19}) under various microscales, while green curve corresponds to that of Eq. (\ref{equ:24}). The data of purple curve and triangle point are obtained from Eq. (8) in \cite{ref21}. It’s obvious that only Rytov variance results $( \eta \to 0$ limit$)$ can be obtained from \cite{ref21}, while we can get general SI results from Eq. (\ref{equ:19}) and reduced Rytov variances from Eq. (\ref{equ:24}), respectively. Besides, from Figure~\ref{fig:02a}, we can find that despite both are Rytov variances, ours and results in \cite{ref21} begin to deviate with the increase of distance. We guess this deviation may be caused by the approximation that eliminates one positive and one negative components in the derivation in \cite{ref21}. However, this deviation seems to vanish when the beam radius is greater, as shown in Figure~\ref{fig:02b}. That is to say, Eq. (8) in \cite{ref21} can be a special case of our exact Eq. (\ref{equ:24}) under wide beam and short distance conditions. Combined Figure~\ref{fig:02a} with Figure~\ref{fig:02b}, we also find the location of red and blue curves occurs a reversal around $L =10 \mathrm{m}$. Meanwhile, take Figure~\ref{fig:02a} as an example, with the microscale increases as $10^{-4} \mathrm{m}$, $10^{-3} \mathrm{m}$ and $10^{-2} \mathrm{m}$, the SI varies with the value of 0.035, 0.038 and 0.006 when $L =15 \mathrm{m}$, respectively. Both phenomena imply that the relation between SI and microscale is nonlinear, which differs with the monotonical relation in \cite{ref09}. This nonlinear relation can be proven by the form of Eq. (\ref{equ:19}). Besides, compared with finite microscale, Rytov variance results overvalue or undervalue turbulence-induced scintillation. So, in order to accurately evaluate SI, the study on the universal case of finite microscale is crucial. It is observed that no matter which curve, the increase of link length leads to a severer turbulence effect, and thus produces a higher SI. At last, due to the narrower beam is less affected by turbulence, so the whole values of SI in Figure~\ref{fig:02a} are lower than Figure~\ref{fig:02b} under the same parameters.\\
\indent In Figure~\ref{fig:03a}, the variation of Rytov variances with Fresnel ratio $\Lambda_0 = 2L/(k W_0^2)$ for three beams is depicted when $W_0 =3 \mathrm{mm}, \eta \to 0$. The data of black curve obtained from Eq. (\ref{equ:24}) and that of pink triangle point obtained from Eq. (8) in \cite{ref21} correspond to Gaussian beam. For plane and spherical waves, blue and purple triangle points correspond to our reduced results from Eq. (\ref{equ:24}) under specific parameter values while red and blue curves correspond to Eqs. (19) and (31) in \cite{ref19}. As we can observe from the comparison, Rytov variances of plane and spherical waves in \cite{ref19} fit well with our results, which indirectly verifies the correctness of our derivation. While the results of Gaussian beam in \cite{ref21} gradually deviate ours, whose reason is identical with Figure~\ref{fig:02a}. Since the beam radius is moderate value $W_0 =3 \mathrm{mm}$, here Eq. (8) in \cite{ref21} cannot be a special case of our exact Eq. (\ref{equ:24}). In regard to Figures~\ref{fig:03b}-\ref{fig:03d}, keeping the microscale $\eta = 10^{-3} \mathrm{m}$, we draw the variation of SI with Fresnel ratio for three beams in case of $W_0 = 1 \mathrm{mm}$, $W_0 = 3 \mathrm{mm}$ and $W_0 = 10 \mathrm{mm}$, respectively. The red, black and blue curves correspond to Eq. (\ref{equ:20}) for plane wave, Eq. (\ref{equ:19}) for Gaussian beam and Eq. (\ref{equ:22}) for spherical wave, respectively. It is shown that the curve of Gaussian beam is closer to that of spherical wave in narrower beam case (Figure~\ref{fig:03b}) while is nearer to plane wave when beam is wider (Figure~\ref{fig:03d}). For a Gaussian beam with moderate beam width, it has unique characteristics which in between plane and spherical waves. Therefore, the study on general SI of Gaussian beam is essential. Comprehensively observing Figure~\ref{fig:03}, we make a conclusion that the Rytov variance and SI of plane wave are the highest than other two beams under the same conditions. It shows that plane wave is greatly affected by oceanic turbulence.\\
\begin{figure*}[t!]
	\centering
	\subfloat[]{
        \includegraphics[width=0.3\linewidth]{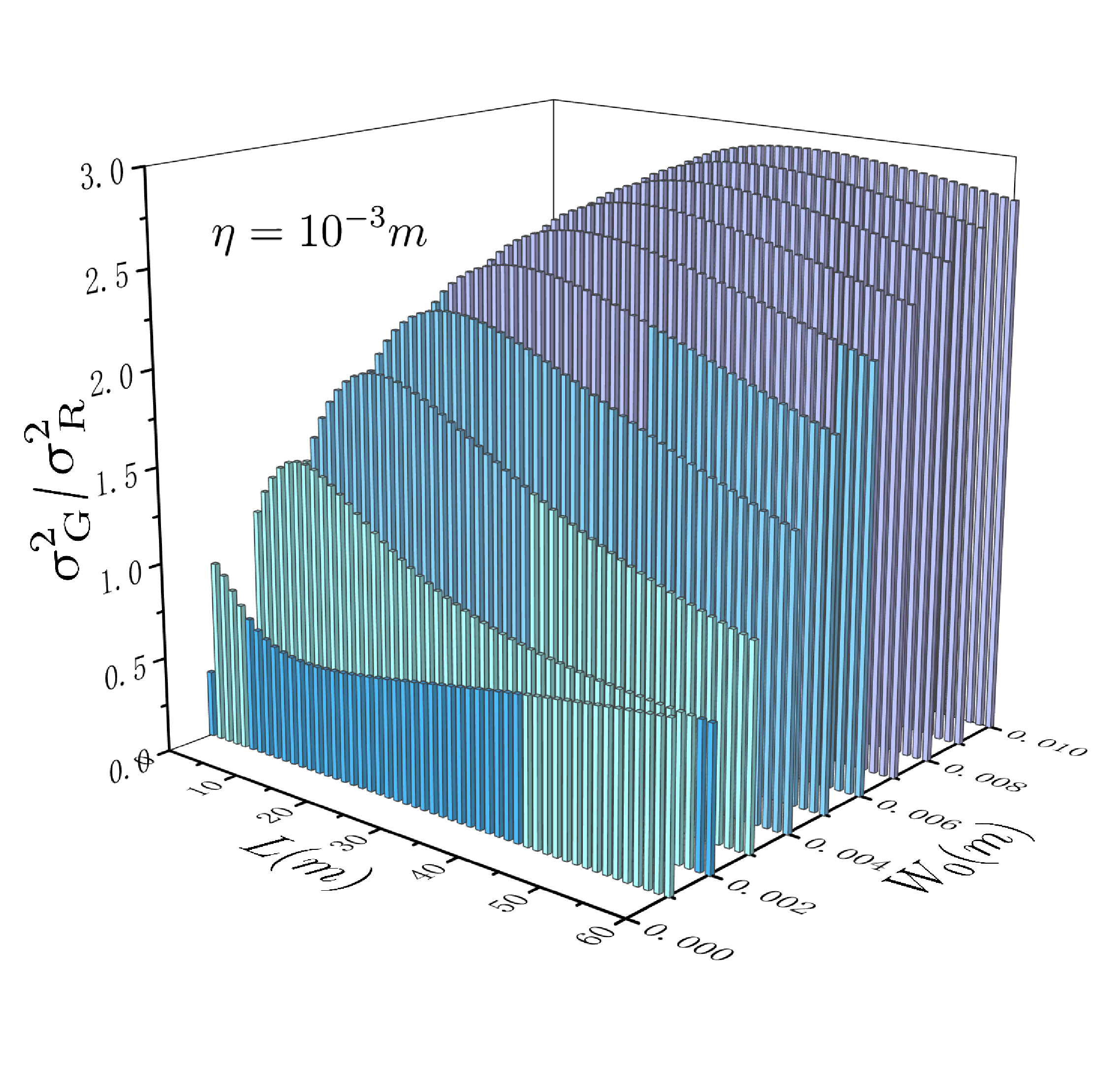} 
        \label{fig:04a} }
	\centering
        \subfloat[]{
        \includegraphics[width=0.3\linewidth]{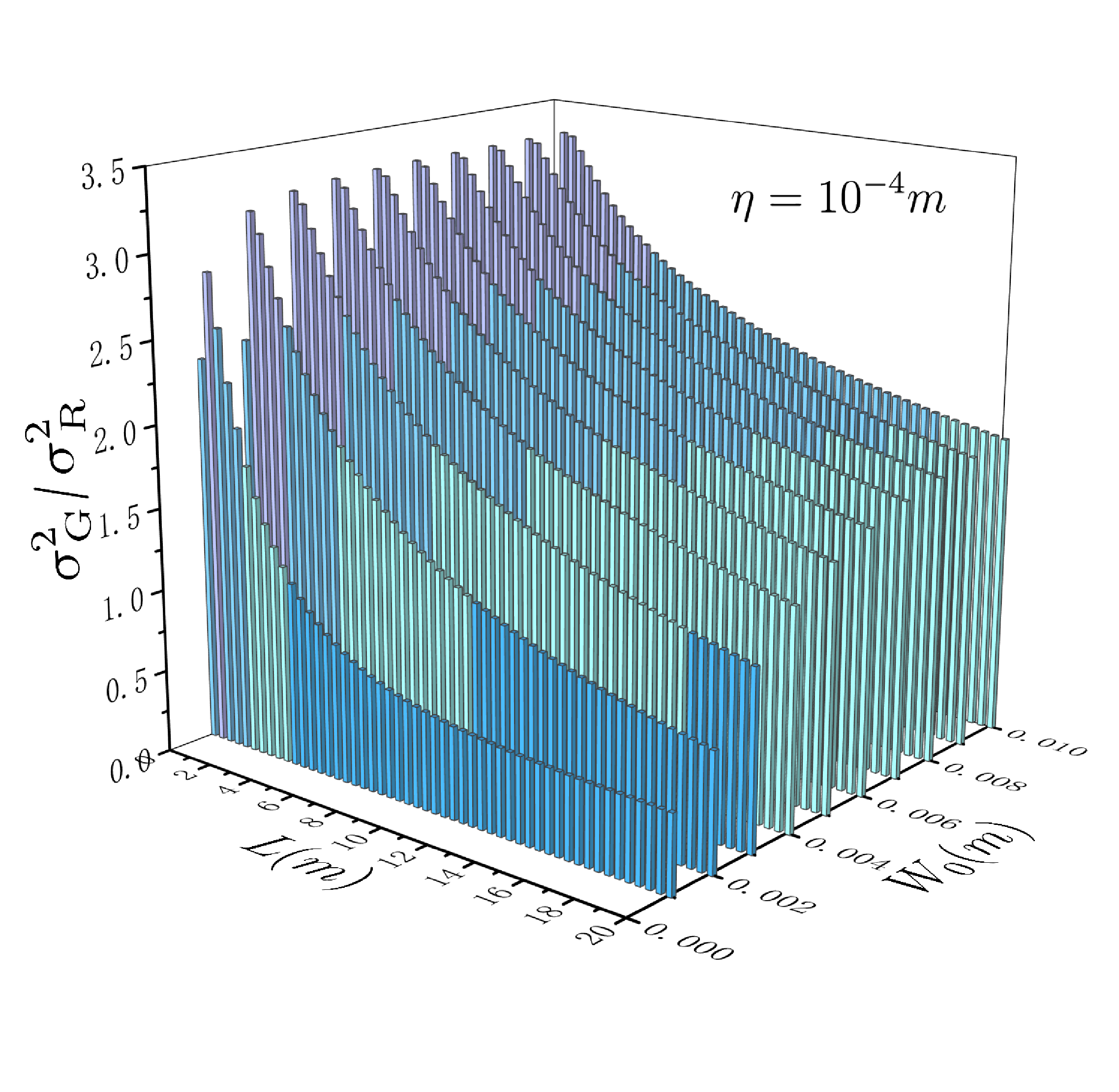} 
        \label{fig:04b} }
        \centering
	\subfloat[]{
        \includegraphics[width=0.3\linewidth]{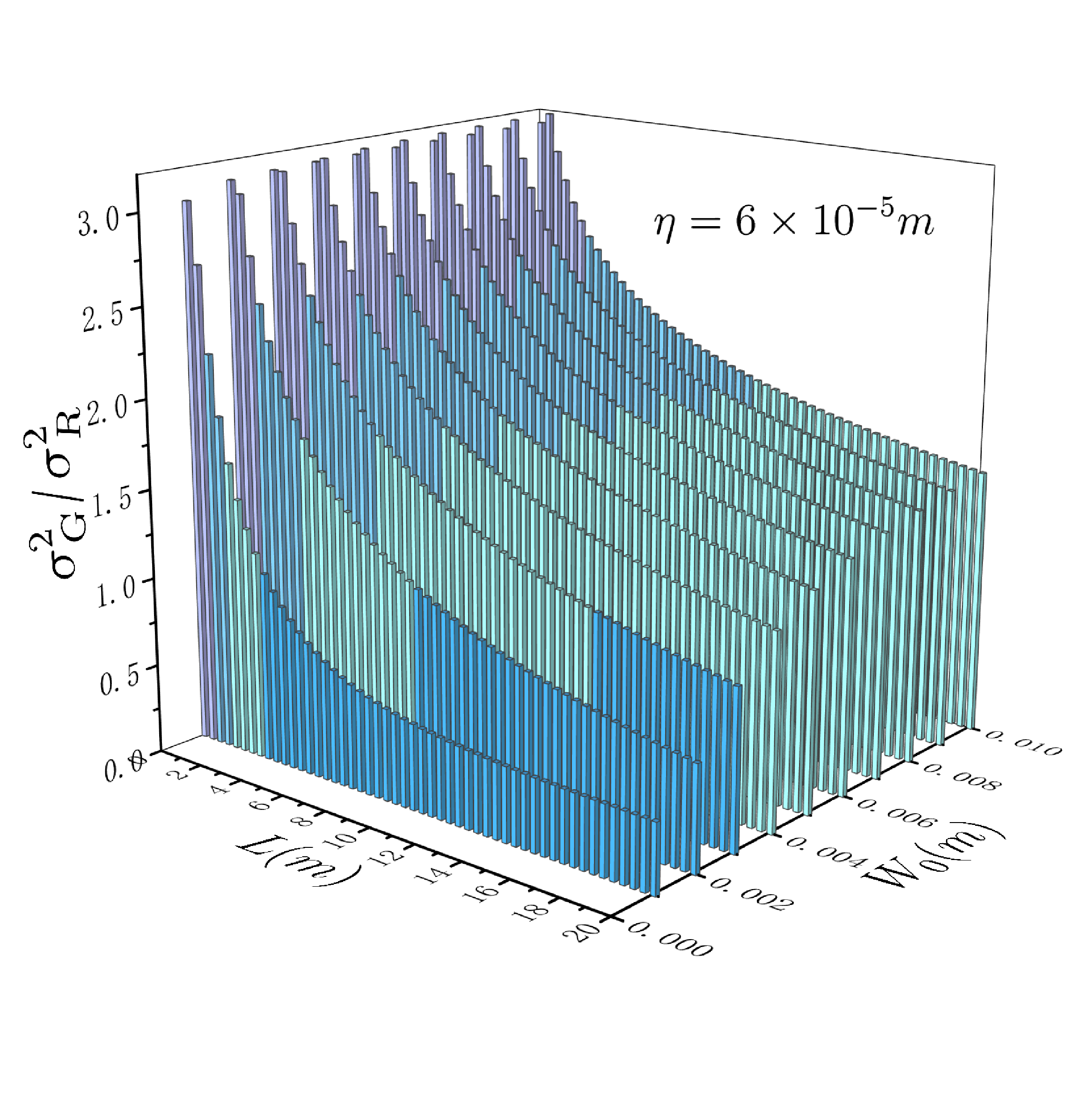} 
        \label{fig:04c} }
	\caption{The ratio of SI to Rytov variance versus link length and beam radius for different microscales, (a) $\eta = 10^{-3}\mathrm{m}$, (b) $\eta = 10^{-4}\mathrm{m}$ and (c) $\eta = 6\times 10^{-5}\mathrm{m}$.}
	\label{fig:04}
\end{figure*}
\begin{figure}[t!]
    \centering
	\subfloat[]{
        \includegraphics[width=0.9\linewidth]{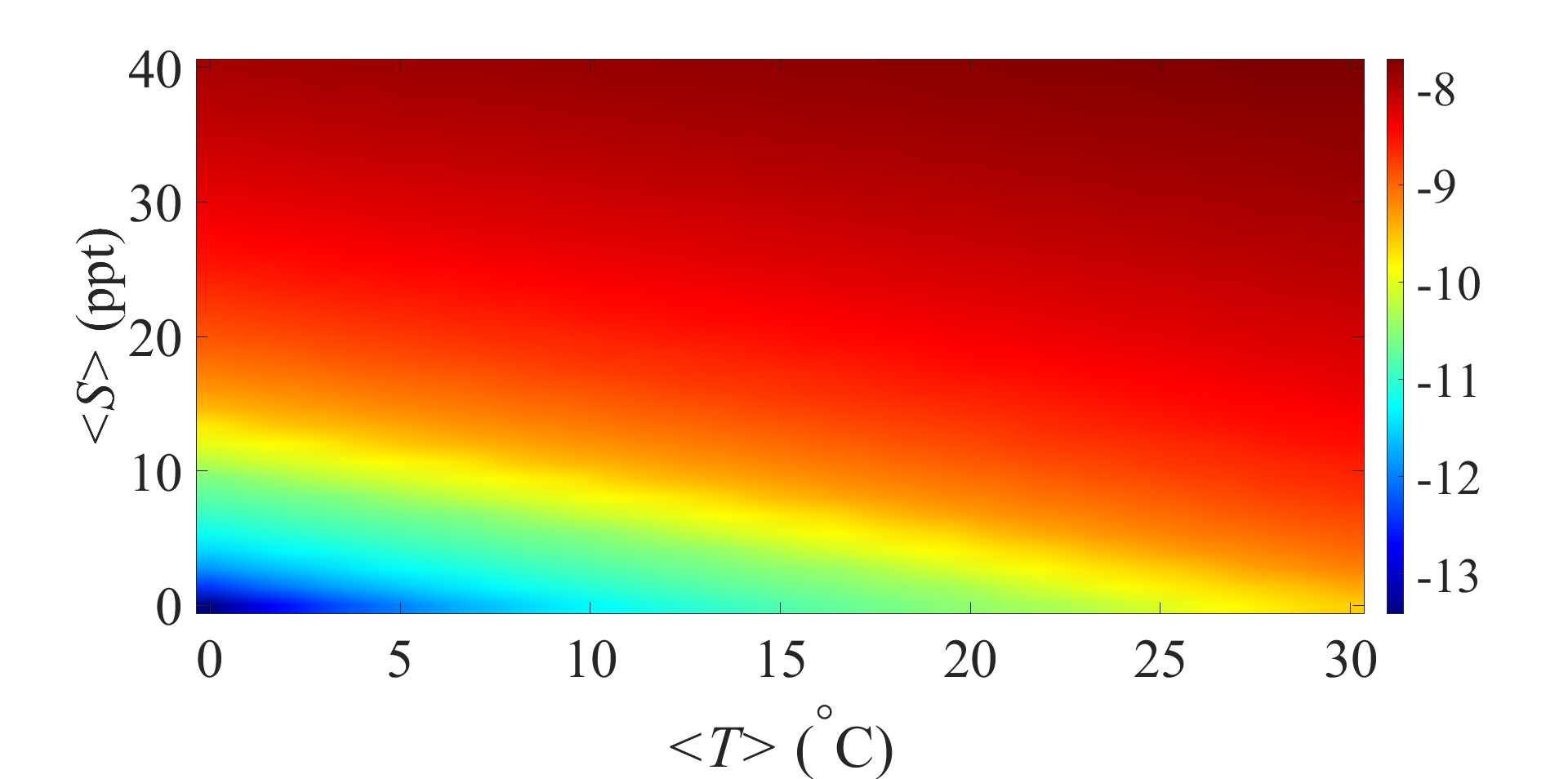} 
        \label{fig:05a} }
        \qquad
	\centering
        \subfloat[]{
        \includegraphics[width=0.9\linewidth]{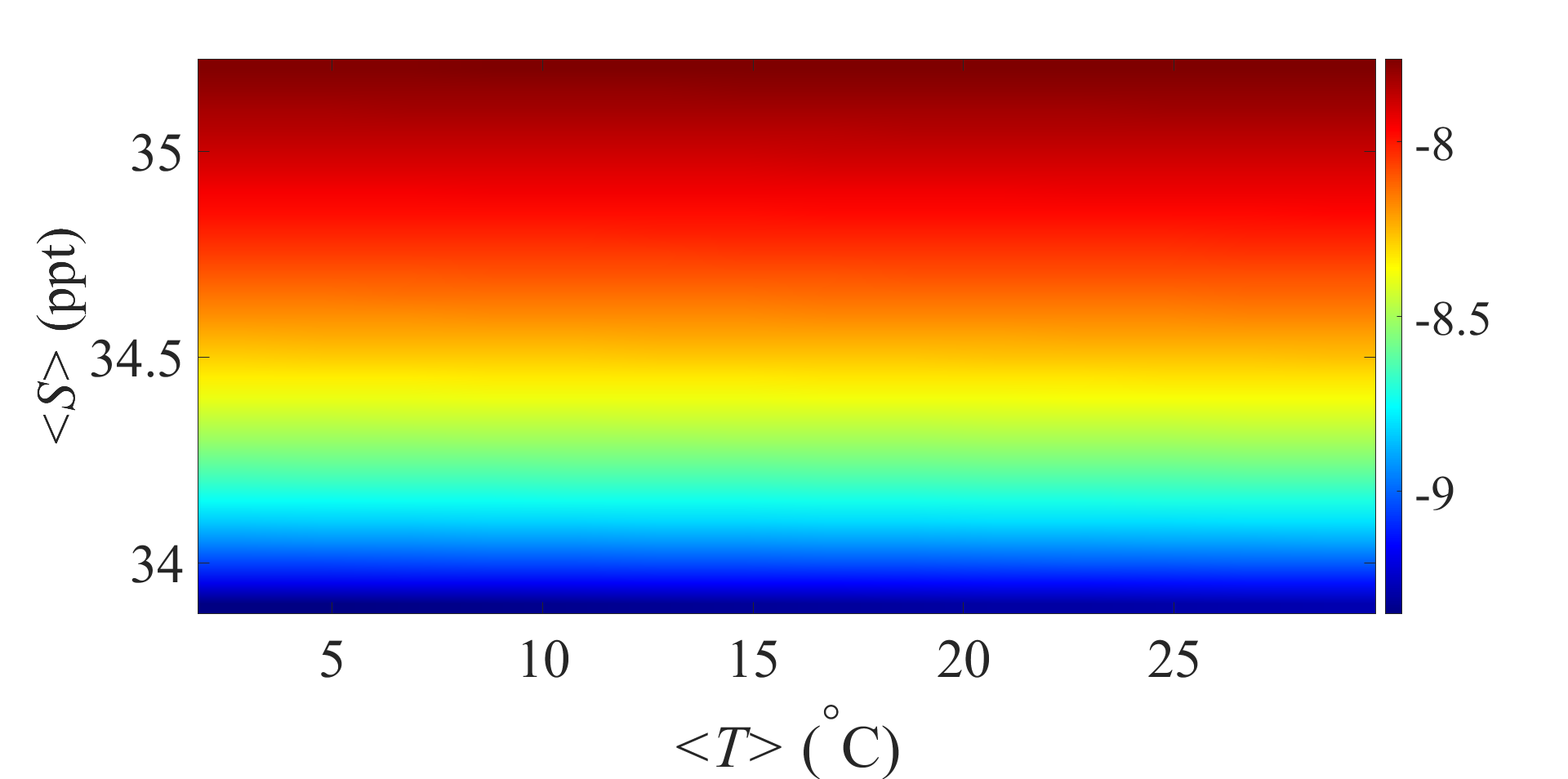} 
        \label{fig:05b} }
        \qquad
        \centering
	\subfloat[]{
        \includegraphics[width=0.9\linewidth]{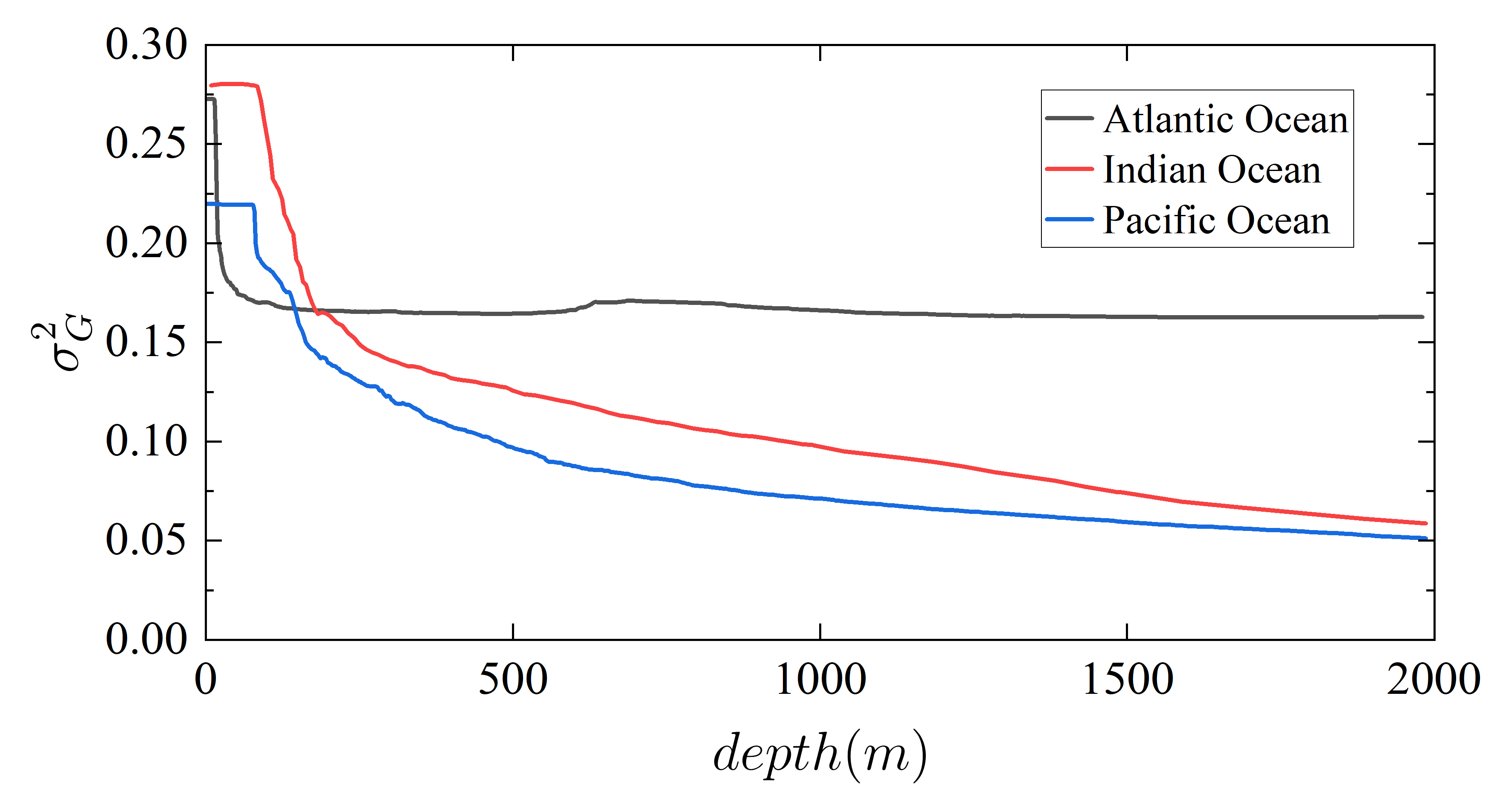} 
        \label{fig:05c} }
	\caption{Density plot of SI -log($\sigma_G^2$)- variation as functions of average temperature and average salinity, (a) wide range of OTOPS, (b) real range of Indian Ocean, and (c) the SI versus depth for different oceans.}
        \label{fig:05}
\end{figure}
\indent The scaled SI $\sigma_G^2 / \sigma_R^2$ for a collimated Gaussian beam is investigated as a function of Fresnel ratio $\Lambda_0 = 2L/(k W_0^2)$ in Figure~\ref{fig:04}, where the two variables (link length $L$ and beam radius $W_0$) have opposite effects on Fresnel ratio. Figures~\ref{fig:04a}-\ref{fig:04c} correspond to $\eta = 10^{-3} \mathrm{m}$, $\eta = 10^{-4} \mathrm{m}$ and $\eta = 6\times 10^{-5} \mathrm{m}$, respectively. Figure~\ref{fig:03} has shown the SI is affected by both link length and beam radius, here we consider combined effects of them. As shown, the scaled SI shows the tendency that first increases and then decreases with the increase of link length. Comparing Figure~\ref{fig:04a} with Figures~\ref{fig:04b} and~\ref{fig:04c}, as the microscale decreases, on the one hand, the peak values appear at shorter distances. On the other hand, the increasing-followed-by-decreasing tendency is already clearly at smaller beam radius. Under the smaller microscales, i.e., in Figures~\ref{fig:04b} and~\ref{fig:04c}, the scaled SI increases with the extension of beam widths while in Figure~\ref{fig:04a}, the monotonic variation relation turns into decreasing followed by increasing when the microscale is larger. This is also reasonable because the forms of Eqs. (\ref{equ:17}) and (\ref{equ:18}) support the nonlinear relation between $W_0$ and SI.\\
\indent We draw the density plot of SI variation $(\log_{}{\sigma_G^2})$ as functions of average temperature $\left \langle T \right \rangle$ and average salinity $\left \langle S \right \rangle$ in Figures~\ref{fig:05a} and~\ref{fig:05b}. Figure~\ref{fig:05a} corresponds to the wide applicable range of the general OTOPS model, i.e., $\left \langle T \right \rangle \in (0,30^{\circ} \mathrm{C})$ and $\left \langle S \right \rangle \in (0,40 \mathrm{ppt})$. However, the $\left \langle T \right \rangle$ and $\left \langle S \right \rangle$ data in Figure~\ref{fig:05b} are from Argo buoy, whose scopes in the range of depth (0, 2000m) in Indian Ocean are $(2.80^{\circ} \mathrm{C},28.71^{\circ} \mathrm{C})$ and $(33.96 \mathrm{ppt},35.11 \mathrm{ppt})$, respectively. It is obvious that the range of change in temperature with depth is much larger than that in salinity. Although both the variation of temperature and salinity can affect the SI by changing corresponding turbulence parameters, salinity-induced turbulence strongly dominates scintillation over temperature-induced \cite{ref08,ref09}. Therefore, the density variation in Figure~\ref{fig:05b} is gentler than Figure~\ref{fig:05a} due to the small range of salinity variation. Additionally, for vertical propagation links, values of temperature and salinity vary with depth, thereby Figure~\ref{fig:05c} depicts the dependence of the SI on depth in different oceans. The results are consistent with the theory, namely, the turbulence on sea surface is more intense, while the turbulence in deep-sea region is gentler. We can also find that the SI of different oceans tend to stabilize at different depths. For Atlantic Ocean, the SI arrives the lowest at near 100m depth. While with regard to Indian Ocean and Pacific Ocean, corresponding depths are more than 1500m. Besides, note that the difference between Figures~\ref{fig:05a},~\ref{fig:05b} and Figure~\ref{fig:05c} is that, for Figures~\ref{fig:05a} and~\ref{fig:05b}, the range of temperature and salinity is determined by scope of application and selected Argo buoy. Then values are taken at uniform intervals from the minimum to the maximum value, that is to say, there may not actually be a specific depth corresponding to a certain set of temperature-salinity values. However, those sets of temperature-salinity values are from specific depths in Figure~\ref{fig:05c}.\\
\indent We study the variation of SI versus link length under different energy dissipation rates in Figure~\ref{fig:06}. Various colors correspond to different energy dissipation rate, characterizing diverse turbulence intensity. As shown, turbulent fluctuations strengthen as the energy dissipation rate decreasing (the direction indicated by the arrow), which leads to a higher SI. Specifically, keeping the distance as $L = 10 \mathrm{m}$, the SI rises from the value of 0.028 to 0.61 with a decrease in $\varepsilon$ from $10^{-1}$ to $ 10^{-5} \mathrm{m^2 s^{-3}}$. It can be also noted that no matter which turbulence intensity is, the longer distance leads to stronger irradiance fluctuations. By observing the gaps between the curves, we can draw a conclusion that longer distance results in larger gaps. This is attributed to the fact that the increase of distance will also aggravate turbulence-induced scintillation. At this point, the results reflect the joint effects of distance and oceanic turbulence parameter $\varepsilon$.\\
\begin{figure}[t!]
    \centering
    \includegraphics[width=0.9\linewidth]{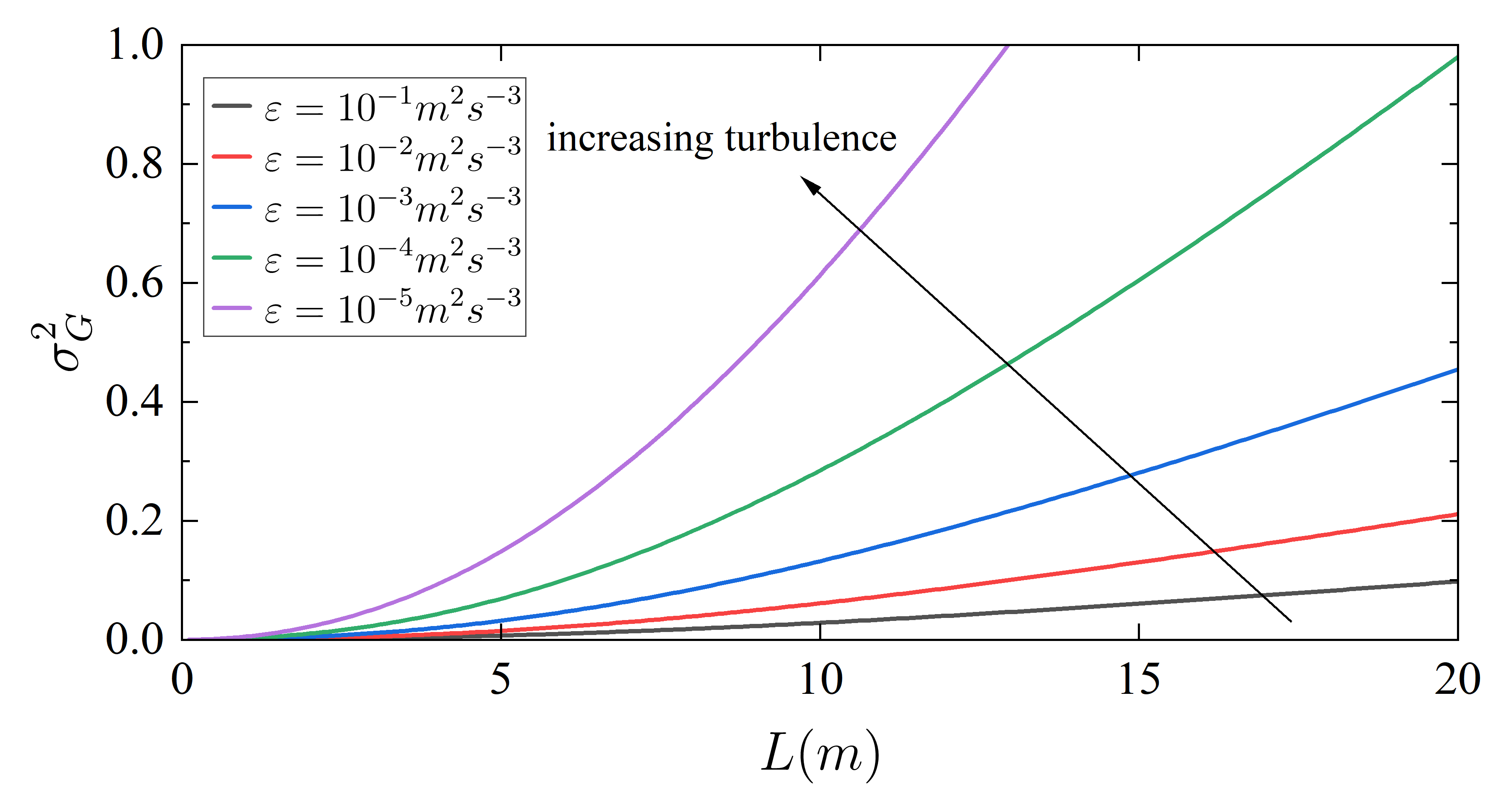}
    \caption{The SI versus link length for different energy dissipation rate $\varepsilon$.}
    \label{fig:06}
\end{figure}
\begin{figure}[t!]
    \centering
    \includegraphics[width=0.9\linewidth]{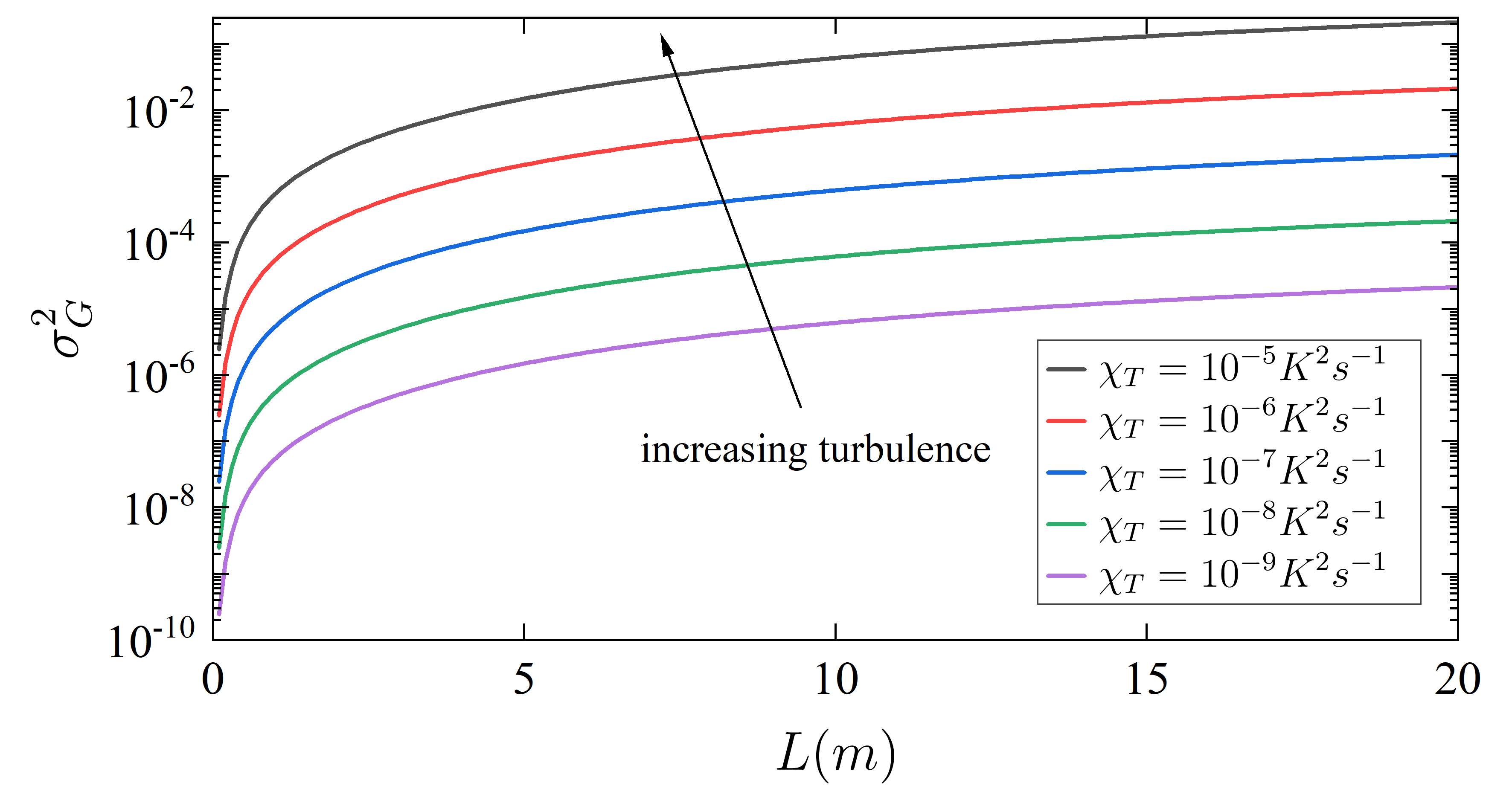}
    \caption{The SI versus link length for different average temperature dissipation rate $\chi_T$.}
    \label{fig:07}
\end{figure}
\indent Figure~\ref{fig:07} depicts the variation of SI versus link length for different average temperature dissipation rate. Average temperature dissipation rate $\chi_T$ is another oceanic turbulence parameter to characterize turbulence intensity likeness to $\varepsilon$. Various colors correspond to different $\chi_T$ and the arrow shows the direction of turbulence enhancement. Apparently, more violent turbulence generates stronger irradiance fluctuations. Specifically, keeping the distance as $L = 10 \mathrm{m}$, the SI rises from the value of $6.13 \times 10^{-6}$ to 0.061 with an increase in $\chi_T$ from $10^{-9}$ to $ 10^{-5} \mathrm{K^2 s^{-1}}$. Analogously, the conclusion that longer distance results in stronger irradiance fluctuations is valid for each curve. Note that we adopt a logarithmic coordinate in y-axis, so the gap between the curves is not identical but significantly increases with the increase of distance. Same as the analysis for Fig. 6, this phenomenon reflects the joint effects of distance and oceanic turbulence parameter $\chi_T$.\\
\subsection{Analysis of the BER Performance}
\begin{figure}[t!]
    \centering
    \includegraphics[width=0.9\linewidth]{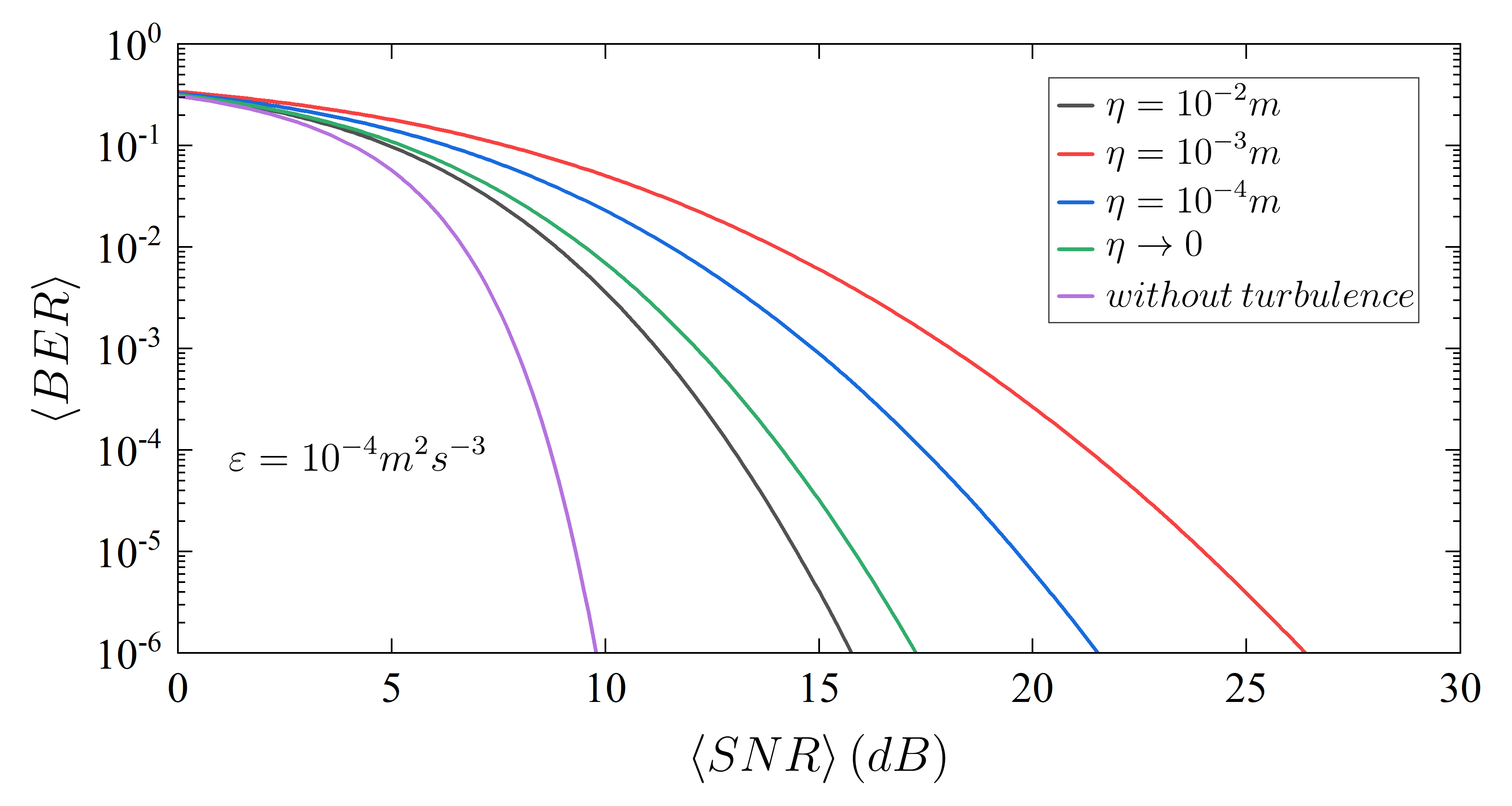}
    \caption{The average BER versus average SNR for various microscales and special case (without turbulence).}
    \label{fig:08}
\end{figure}

\indent The average BER variation of UWOC systems depending on the average SNR is plotted in Figure~\ref{fig:08}. Purple curve corresponds to special case without turbulence, other four colors’ curves correspond to various microscales, where the green curve represents limit case $\eta \to 0$. Firstly, it is obvious that the average BER monotonically decreases with the increase of average SNR, no matter which condition. Next, compared with the case of turbulence existing, results without turbulence shows a better BER performance and needs lower $\left \langle SNR \right \rangle$ to meet the same $\left \langle BER \right \rangle$. Furthermore, as we analyzed earlier, the influence of microscale $\eta$ on SI is nonlinear, thus its influence on average BER is also nonlinear. Specifically, if we set $\left \langle BER \right \rangle$ below $10^{-6}$ is acceptable, then $\left \langle SNR \right \rangle$ to meet requirements increases nearly 11 dB with a variation in the microscale from $10^{-3}$ to $10^{-2} \mathrm{m}$. Moreover, we can note that the curve of limit case $\eta \to 0$ is between curve of $\eta = 10^{-2} \mathrm{m}$ and $\eta = 10^{-3} \mathrm{m} (\eta = 10^{-4} \mathrm{m})$. It proves again that Rytov variance results (limit case $\eta \to 0$) will overvalue or undervalue the associated SI and BER, suggesting the necessity of our universal work containing finite microscale.\\
\begin{figure}[t!]
    \centering
    \includegraphics[width=0.9\linewidth]{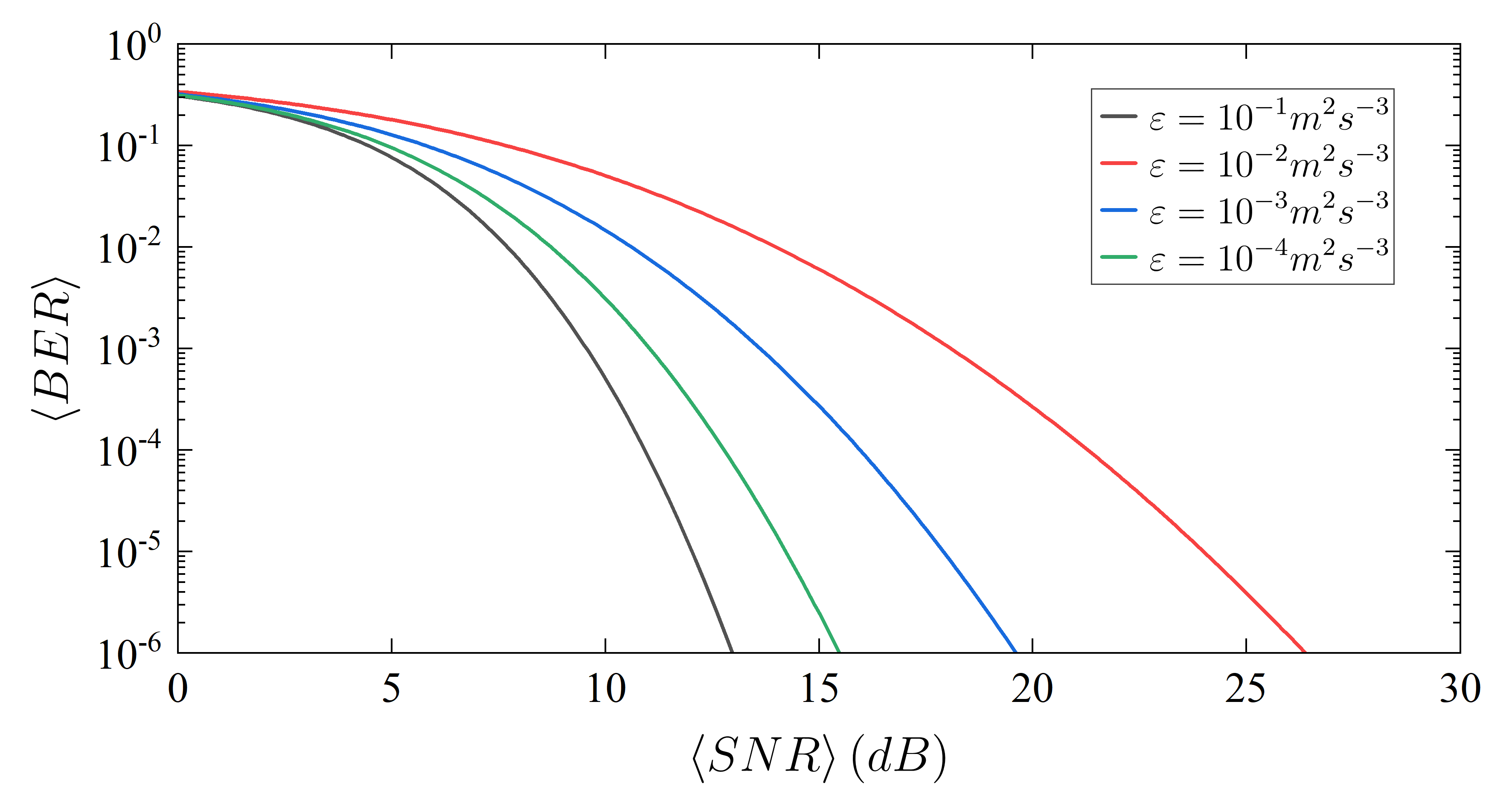}
    \caption{The average BER versus average SNR for various energy dissipation rate $\varepsilon$.}
    \label{fig:09}
\end{figure}
\begin{figure}[t!]
    \centering
    \includegraphics[width=0.9\linewidth]{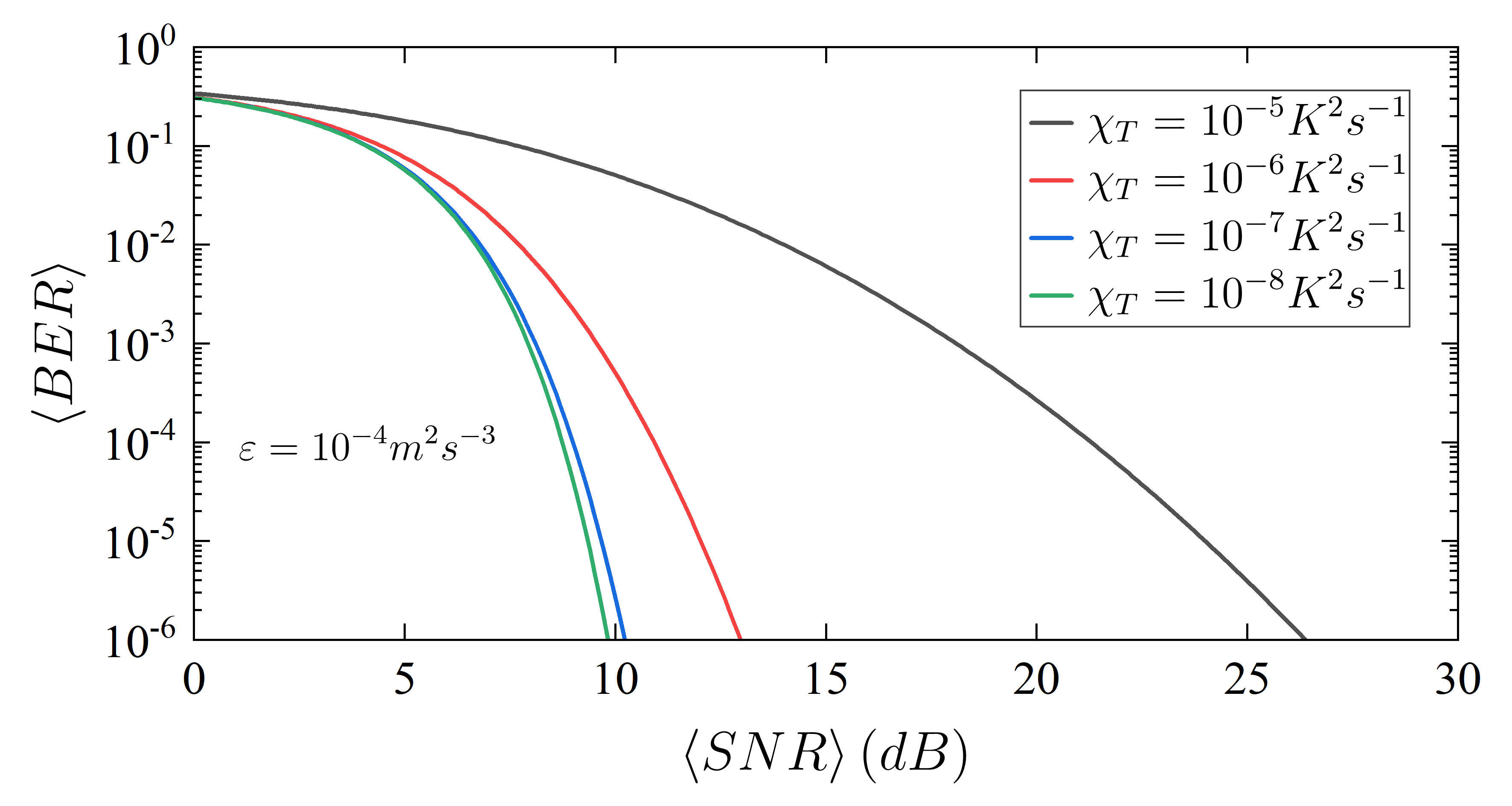}
    \caption{The average BER versus average SNR for various average temperature dissipation rate $\chi_T$.}
    \label{fig:10}
\end{figure}
\indent The average BER performance of UWOC systems for different energy dissipation rates is shown in Figure~\ref{fig:09}. Various colors correspond to different energy dissipation rate $\varepsilon$. The identical conclusion that $\left \langle BER \right \rangle$ monotonically decreases with the increase of $\left \langle SNR \right \rangle$ still holds in this figure. On the one hand, under the same $\left \langle SNR \right \rangle$ at the receiver, the more violent turbulence characterized by lower $\varepsilon$ produces a higher $\left \langle BER \right \rangle$. On the other hand, to drop $\left \langle BER \right \rangle$ to an acceptable threshold (generally select $10^{-6}$), a larger $\left \langle SNR \right \rangle$ is needed under a stronger fluctuation. Specifically, keeping the average BER as $10^{-6}$, the average SNR needed rises from $15.4 \mathrm{dB}$ to $26.3 \mathrm{dB}$ with an increase in $\varepsilon$ from $10^{-4}$ to $10^{-2} \mathrm{m^2 s^{-3}}$.\\
\indent Figure~\ref{fig:10} depicts the variation of average BER versus average SNR for different average temperature dissipation rate. Various colors correspond to different average temperature dissipation rate $\chi_T$. It is obvious that increasing the average SNR can improve BER performance of UWOC systems markedly. With $\left \langle SNR \right \rangle$ increases as $5 \mathrm{dB}$, $15 \mathrm{dB}$, and $25 \mathrm{dB}$, $\left \langle BER \right \rangle$ drops with the value of 0.18, $6.07 \times 10^{-3}$ and $3.93 \times 10^{-6}$ when $\chi_T = 10^{-5} \mathrm{K^2 s^{-1}}$, respectively. Moreover, under the same $\left \langle SNR \right \rangle$ at the receiver, the more violent turbulence characterized by larger $\chi_T$ appears a worse BER performance. Another conclusion is that with the increase of $\chi_T$, $\left \langle SNR \right \rangle$ to meet the threshold of $\left \langle BER \right \rangle$ $10^{-6}$ increases accordingly. Specifically, $\left \langle SNR \right \rangle$ rises from the value of $9.8 \mathrm{dB}$ to $26.3 \mathrm{dB}$ with an increase in $\chi_T$ from $10^{-8}$ to $10^{-5} \mathrm{K^2 s^{-1}}$.\\
\section{Conclusion} \label{sec:conclusion}
In this paper, we make a detailed derivation of the closed-form expression for the SI of Gaussian beam based on the general OTOPS and Rytov theory. On the basis of derived SI, we analyze the effects of different parameters on the SI and the BER performance. Besides including the previous results in \cite{ref19,ref21}, our work extends the applicable conditions to general case the microscale being non-zero. Furthermore, numerical results demonstrate that the relationship between microscale and SI is nonlinear, which implies that the Rytov variance results (zero microscale limit) may either overestimate or underestimate the SI and BER. Thus, we suggest that the estimations of SI and BER for real oceans should be within a range rather than a fixed value. In addition, we use temperature and salinity data from Argo buoy to study the dependence of the SI on depth. We expect that our work will benefit to build a more exact channel model and help design a more robust and reliable UWOC system in underwater application scenarios.
\section*{Funding}
This work is supported by National Key Research and Development Program of China (2022YFC2808101) and National Natural Science Foundation of China (61505155).

\end{document}